\documentclass[12pt]{article}
\usepackage{latexsym}
\usepackage{amsmath}
\usepackage{amssymb}
\usepackage{relsize}
\usepackage{geometry}
\def\beqn{\begin{eqnarray}}
\def\eeqn{\end{eqnarray}}

\def\beq{\begin{equation}}
\def\eeq{\end{equation}}
\def\ba{\beq\new\begin{array}{c}}
\def\ea{\end{array}\eeq}

\def\Tr{{\rm Tr}}
\newcommand{\gsim}{\lower.7ex\hbox{$
\;\stackrel{\textstyle>}{\sim}\;$}}
\newcommand{\lsim}{\lower.7ex\hbox{$
\;\stackrel{\textstyle<}{\sim}\;$}}

\newcommand{\ntwo}{${\mathcal N}=2$ }
\newcommand{\ntwon}{${\mathcal N}=2$}
\newcommand{\ntwot}{${\mathcal N}= \left(2,2\right) $ }
\newcommand{\ntwoo}{${\mathcal N}= \left(0,2\right) $ }
\newcommand{\none}{${\mathcal N}=1$ }
\newcommand{\nonen}{${\mathcal N}=1$}

\newcommand{\p}{\partial}
\newcommand{\wt}{\widetilde}
\newcommand{\ov}{\overline}
\newcommand{\mc}[1]{\mathcal{#1}}
\newcommand{\md}{\mathcal{D}}

\newcommand{\lgr}{\left\lgroup}
\newcommand{\rgr}{\right\rgroup}

\def\slashed#1{\setbox0=\hbox{$#1$}             
   \dimen0=\wd0                                 
   \setbox1=\hbox{/} \dimen1=\wd1               
   \ifdim\dimen0>\dimen1                        
      \rlap{\hbox to \dimen0{\hfil/\hfil}}      
      #1                                        
   \else                                        
      \rlap{\hbox to \dimen1{\hfil$#1$\hfil}}   
      /                                         
   \fi}                                        %

\newcommand{\LN}{\Lambda_\text{SU($N$)}}
\newcommand{\sunu}{{\rm SU($N$) $\times$ U(1)} }
\newcommand{\sunun}{{\rm SU($N$) $\times$ U(1)}}
\def\cfl {$\text{SU($N$)}_{\rm C+F}$ }
\newcommand{\mUp}{m_{\rm U(1)}^{+}}
\newcommand{\mUm}{m_{\rm U(1)}^{-}}
\newcommand{\mNp}{m_\text{SU($N$)}^{+}}
\newcommand{\mNm}{m_\text{SU($N$)}^{-}}
\newcommand{\AU}{\mc{A}^{\rm U(1)}}
\newcommand{\AN}{\mc{A}^\text{SU($N$)}}
\newcommand{\aU}{a^{\rm U(1)}}
\newcommand{\aN}{a^\text{SU($N$)}}
\newcommand{\baU}{\ov{a}{}^{\rm U(1)}}
\newcommand{\baN}{\ov{a}{}^\text{SU($N$)}}
\newcommand{\lU}{\lambda^{\rm U(1)}}
\newcommand{\lN}{\lambda^\text{SU($N$)}}
\newcommand{\bxir}{\ov{\xi}{}_R}
\newcommand{\bxil}{\ov{\xi}{}_L}
\newcommand{\xir}{\xi_R}
\newcommand{\xil}{\xi_L}
\newcommand{\bzl}{\ov{\zeta}{}_L}
\newcommand{\bzr}{\ov{\zeta}{}_R}
\newcommand{\zr}{\zeta_R}
\newcommand{\zl}{\zeta_L}
\newcommand{\nbar}{\ov{n}}

\newcommand{\loU}{\lambda_0^{\rm U(1)}}
\newcommand{\llU}{\lambda_1^{\rm U(1)}}
\newcommand{\loN}{\lambda_0^\text{SU($N$)}}
\newcommand{\llN}{\lambda_1^\text{SU($N$)}}
\newcommand{\poU}{\psi_0^{\rm U(1)}}
\newcommand{\plU}{\psi_1^{\rm U(1)}}
\newcommand{\poN}{\psi_0^\text{SU($N$)}}
\newcommand{\plN}{\psi_1^\text{SU($N$)}}

\newcommand{\CPC}{CP($N-1$)$\times$C }
\newcommand{\CPCn}{CP($N-1$)$\times$C}

\newcommand{\MN}{M_\text{SU($N$)}}
\newcommand{\MU}{M_{\rm U(1)}}

\newcommand{\tgamma}{\wt{\gamma}}

\begin{document}

%
%

%
%

\begin{titlepage}

\begin{flushright}
FTPI-MINN-09/03 UMN-TH-2734/09\\
January 28, 2009
\end{flushright}

\begin{center}

{\Large \bf   Description of the Heterotic String Solutions
\\[2mm]
 in \boldmath{${\rm U}(N)$} SQCD }
\end{center}

\begin{center}
{\bf P. A. Bolokhov$^{a,b}$, M.~Shifman$^{c}$, and \bf A.~Yung$^{c,d,e}$}
\end {center}
\vspace{0.3cm}
\begin{center}

$^a${\it Physics and Astronomy Department, University of Pittsburgh, Pittsburgh, Pennsylvania, 15260, USA}\\
$^b${\it Theoretical Physics Department, St.Petersburg State University, Ulyanovskaya~1, 
	 Peterhof, St.Petersburg, 198504, Russia}\\
$^c${\it  William I. Fine Theoretical Physics Institute,
University of Minnesota,
Minneapolis, MN 55455, USA}\\
$^{d}${\it Petersburg Nuclear Physics Institute, Gatchina, St. Petersburg
188300, Russia}\\
$^e${\it Institute of Theoretical and Experimental Physics, Moscow
117259, Russia}
\end{center}

\begin{abstract}

We continue the  study of heterotic non-Abelian BPS-saturated flux tubes (strings).
Previously, such solutions were obtained \cite{SYhet} in a particular U(2) gauge theory:
 \ntwo super\-symmetric QCD deformed by
superpotential terms of a special type breaking
\ntwo supersymmetry down to \none$\!\!$. Here we generalize the previous results
to U$(N)$ gauge theories.
As was suggested by Edalati and Tong \cite{Edalati}, the string world sheet theory
is a heterotic \ntwoo 
sigma model, with the CP$(N-1)$ target space for bosonic fields and an extra 
right-handed fermion which couples to the fermion fields of the
\ntwot CP$(N-1)$ model. We derive the heterotic  \ntwoo world sheet
model directly from the U$(N)$ bulk theory. Parameters of the bulk theory are related to those of the world sheet theory. Qualitatively this relation turns out to be  the same as in the U(2) case.

\end{abstract}

\end{titlepage}

\section{Introduction}

The simplest model supporting non-Abelian flux tubes (strings)
has the gauge group
U$(N)$, with the U(1) Fayet--Iliopoulos (FI) term,
and $N$ flavors of quarks ($N$ hypermultiplets in the fundamental representation;
for a review see \cite{SYrev}).
A crucial feature of  non-Abelian strings
is the presence of orientational (and superorientational) moduli associated with
rotations of their color fluxes inside a non-Abelian group, in addition to
``standard" translational and supertranslational moduli
\cite{HT1,ABEKY,SYmon,HT2}, see also the review papers \cite{Trev,SYrev,Jrev,Trev2}. 
If \ntwo supersymmetry is maintained in the bulk,
the low-energy theory on the
string world sheet is split into two disconnected parts:
a free theory for (super)translational moduli and a nontrivial part, a theory of
interacting (super)orientational moduli described by the CP$(N-1)$ model.
 The above split of the moduli space
 is completely fixed by the fact
that the basic bulk theory has eight supercharges, and the string under consideration
is 1/2 BPS (classically). 

If \ntwo bulk theory is deformed by mass terms of the adjoint fields
breaking \ntwo down to \nonen, the situation drastically changes:
two of the four former supertranslational modes
become coupled to two superorientational modes \cite{Edalati}.
As a result, the world sheet theory is deformed too.
Instead of the well-studied ${\mathcal N}=(2,2)$ CP$(N-1)$ model we now have
a heterotic \ntwoo 
sigma model, with the CP$(N-1)$ target space for bosonic fields and an extra 
right-handed fermion which couples to the fermion fields of the
\ntwot CP$(N-1)$ model in a special way. In the previous work \cite{SYhet}
the heterotic world sheet model was derived
in a particular case of the U(2) bulk theory. 
Our present task is to extend this derivation
to U$(N)$ theories with arbitrary $N$.

To this end a significant work must be done: all previously undetermined fermion zero modes  
need to be found. We do this job in the limit of small and large coefficients in front of the
\ntwo breaking terms. Having these explicit expressions it is straightforward
to relate the  heterotic \ntwoo 
sigma model parameters with those of the bulk theory.

The paper is organized as follows. In Sect.~\ref{MT} we review the
basic \ntwo bulk theory and introduce a deformation that breaks supersymmetry down to \nonen. This deformation is of a special form: it gives masses to all adjoint
fields.
Then we outline the non-Abelian string solution
in the U$(N)$ theory (Sect.~\ref{vortex}). Starting from the unbroken \ntwo bulk theory
with the U$(N)$ gauge group (Sect.~\ref{zeromodes}) and moving towards
the deformed \none theory we determine all fermion zero modes (Sect.~\ref{WS}).
In Sect.~\ref{BC} we derive the relation between the bulk and world sheet parameters.
In Sect.~\ref{FT} we discuss the geometric formulation of the heterotic CP$(N-1)$ model.
 In Sect.~\ref{largeN} we give  a brief review
of  physics of the \ntwoo CP$(N-1)$ model, while Sect.~9 presents our conclusions. 
Our notation is summarized and explained in Appendix.


\section{Microscopic Theory}
\label{MT}
\setcounter{equation}{0}

In this section we introduce the theory in the bulk, and review its perturbative mass spectrum.
The starting theory is $\mc{N}=2$ SQCD with the gauge group \sunun. 
Its matter sector consists of $ N_f = N $ flavors of quark hypermultiplets,
both in fundamental and antifundamental representations, as necessary for $ \mc{N}=2 $ supersymmetry.
For the string solutions to exist, we add  the Fayet--Iliopoulos (FI) $D$-term, which causes quark
condensation.
The superpotential of the theory 
\beq
\label{ntwo}
	\mc{W}_{\mc{N}=2} ~~=~~  \sqrt{2}\, \Bigl\{ 
					\wt{q}{}_A \AU q^A ~+~
					\wt{q}{}_A \mc{A}^a T^a q^A \Bigr\}  ~+~
				m_A\, \wt{q}{}_A q^A
\eeq
includes the quark multiplets $ q^A $ and $ \wt{q}{}_A $ ($A = 1, ..., N$), and the adjoint matter multiplets
$ \AU $ and $ \AN = \mc{A}^a T^a $ which are the $ \mc{N}=2 $ superpartners of the U(1) and SU($N$) 
gauge multiplets.

To break supersymmetry to \nonen, we introduce mass terms for the adjoint matter fields
\beq
\label{none}
	\mc{W}_{\mc{N}=1} ~~=~~ \sqrt{2N}\,\mu_1 \left(\mc{A}^{\rm U(1)}\right)^2  ~~+~~
				\frac{\mu_2}{2} \left( \mc{A}^a \right)^2 ~.
\eeq
Numerical factors here were chosen for normalization purposes. 
The masses $ \mu_1 $ and $ \mu_2 $ lift the adjoints above their gauge superpartners and,
in  this way, break \ntwo
supersymmetry.
Although the parameters $ \mu_1 $ and $ \mu_2 $ are generally speaking different, we will later assume that
they are connected by a particular relation.
The latter is, of course, not essential, as our goal is the limit of very large $ \mu_1$ and $\mu_2$. 

The theory broken by \eqref{none} admits 1/2 BPS-saturated vortex solutions, if the mass parameters
are set to zero \cite{SYrev,SYnone,Edalati},
\beq
\label{qmasses}
	m_1 ~~=~~ m_2 ~~=~~ \dots ~~=~~ m_N ~~=~~ 0~.
\eeq
The bosonic part of the theory takes the form
\begin{align}
\label{theory}
	S_{\rm bos} ~~=~~ & \int d^4 x 
		\lgr
			\frac{1}{2g_2^2}\Tr \left(F_{\mu\nu}^\text{SU($N$)}\right)^2  ~+~
			\frac{1}{g_1^2} \left(F_{\mu\nu}^{\rm U(1)}\right)^2 ~+~ 
			\right. 
			\\[2mm]
\notag
		&
			\phantom{int d^4 x \lgr\right.}
			\frac{2}{g_2^2}\Tr \left|\nabla_\mu \aN \right|^2   ~+~
			\frac{4}{g_1^2} \left|\p_\mu \aU \right|^2
			~+~
			\left| \nabla_\mu q^A \right|^2 ~+~ \left|\nabla_\mu \ov{\wt{q}}{}^A \right|^2 
			~+~
			\\[2mm]
\notag
		&
			\phantom{int d^4 x \lgr\right.}
		\left.
			V(q^A, \wt{q}{}_A, \aN, \aU)
		\rgr .
\end{align}
Here $ \nabla_\mu $ denotes the covariant derivative in the appropriate representation
\begin{align*}
	\nabla_\mu^{\rm adj} & ~~=~~ \p_\mu  ~-~ i\, [ A_\mu^a T^a, \;\cdot\;]~, \\[3mm]
	\nabla_\mu^{\rm fund} & ~~=~~ \p_\mu ~-~ i\,A^{\rm U(1)}_\mu ~-~ i\, A_\mu^a T^a~,
\end{align*}
with the SU($N$) generators normalized as
\[
	\Tr \left( T^a T^b \right) ~=~ (1/2) \, \delta^{ab}~.
\]

The potential is given by the  sum of the $ D $ and $ F $ terms,
\begin{align}
\notag
	& V(q^A, \wt{q}{}_A, \aN, \aU) ~~=~~ 
	\\[3mm]
\notag
	&\qquad\quad ~~=~~
			\frac{g_2^2}{2} \left( \frac{1}{g_2^2}\,f^{abc}\ov{a}{}^b a^c 
				~+~ \ov{q}{}_A\, T^a q^A ~-~ \wt{q}{}_A\, T^a \ov{\wt{q}}{}^A \right)^2 
	\\[3mm]
\label{V}
	&\qquad\quad ~~+~~
		\frac{g_1^2}{8}\, (\ov{q}{}_A q^A ~-~ \wt{q}{}_A \ov{\wt{q}}{}^A ~-~ N\xi )^2
	\\[3mm]
\notag
	&\qquad\quad ~~+~~
		2g_2^2\, \Bigl| \wt{q}{}_A\,T^a q^A ~+~ 
			\frac{1}{\sqrt{2}}\, \frac{\p\mc{W}_{\mc{N}=1}}{\p a^a} \Bigr|^2
	~+~
	\frac{g_1^2}{2}\, \Bigl| \wt{q}{}_A q^A ~+~ 
			\frac{1}{\sqrt{2}}\, \frac{\p\mc{W}_{\mc{N}=1}}{\p\aU} \Bigr|^2
	\\[3mm]
\notag
	&\qquad\quad ~~+~~
	2 \sum_{A=1}^{N} \Biggl\{  
		\left| \left( \aU ~+~ \frac{m_A}{\sqrt{2}} ~+~ a^a T^a \right) q^A \right|^2  ~+~
	\\[3mm]
\notag
	&\phantom{\qquad\quad ~~+~~ 2 \sum_{A=1}^{N} \Biggl\{  }
		\left| \left( \aU ~+~ \frac{m_A}{\sqrt{2}} ~+~ a^a T^a \right) \ov{\wt{q}}{}^A \right|^2  
			\Biggr\}~,
\end{align}
	where summation is implied over the repeated flavor indices $A$.
	Here $\xi$ is the parameter of the Fayet--Iliopoulos $ D $-term. 

	The perturbative spectrum of this model was derived in detail in \cite{SYrev}; here
	we review just some of relevant results. 
	The role of the Fayet--Iliopoulos term is to trigger spontaneous breaking of the gauge 
	symmetry.
	The vacuum expectation values (VEVs) of the scalar quarks can be chosen in the 
	color-flavor locked form
\begin{align}
\notag
&
	\langle q^{kA} \rangle ~=~ \sqrt{\xi} 
		\begin{pmatrix}
			 1  &   0  &  ... \\
			... &  ... &  ... \\
			... &   0  &  1 
		\end{pmatrix} ~,
	\qquad\qquad 
	\langle \ov{\wt{q}}{}^{kA} \rangle ~=~ 0~,
	\\
\label{qVEV}
&
	\qquad\qquad  k~=~ 1,...\, N~, \qquad  A ~=~ 1,...\, N~,
\end{align}
	while the VEVs of the adjoint fields vanish
\beq
\label{aVEV}
	\langle \aN \rangle  ~~=~~ 0~, \qquad\qquad  \langle \aU \rangle ~~=~~ 0~.
\eeq
The color-flavor locking of the quark VEVs implies that the global \cfl 
	symmetry is unbroken in the vacuum.
	Much in the same way as in \ntwo SQCD, this symmetry leads to the emergence of the orientational
	zero modes of the $ Z_N $ strings.

The values of the parameters in \eqref{V} are chosen in such a way that the adjoint VEVs vanish, 
	and, therefore, the VEVs \eqref{qVEV} and \eqref{aVEV} do not depend on the supersymmetry breaking
	parameters $ \mu_1 $ and $ \mu_2 $.
	In particular, the same pattern of the symmetry breaking will be observed all the way up to
	very large $ \mu_1 $ and $ \mu_2 $, where the adjoints decouple.
	As in \ntwo SQCD, we assume $ \sqrt{\xi} ~\gg~ \LN $ to ensure weak coupling.

	Now, since both U(1) and SU($N$) gauge groups are broken by squark condensation, all gauge bosons
	become massive, with masses 
\beq
\label{MN}
	\MN ~~=~~ g_2\sqrt{\xi}
\eeq
	and
\beq
\label{MU}
	\MU ~~=~~ g_1 \sqrt{\frac{N}{2}} \xi~.
\eeq
	To obtain the scalar boson masses one needs to expand the potential \eqref{V} near 
	the vacuum \eqref{qVEV}, \eqref{aVEV} and diagonalize the corresponding mass matrix.
	Then, $ N^2 $ components of $2N^2$ (real) component scalar field $ q^{kA} $ are eaten by the
	Higgs mechanism for the U(1) and SU($N$) gauge groups, respectively. 
	Other $ N^2 $ components are split as follows: one component acquires mass $ \MU $.
	It becomes the scalar component of a massive \none vector U(1) gauge multiplet.
	Moreover, $ N^2 - 1 $ components acquire masses $ \MN $ and become superpartners of the
	SU($N$) gauge bosons in \none  massive gauge supermultiplets.
	
	Other $ 4 N^2 $ real scalar components of fields $ \wt{q}{}_{Ak} $, $\aN$ and $\aU$ produce the
	following states: 
	two states acquire masses
\beq
\label{mUp}
	\mUp ~~=~~ g_1 \sqrt{\frac{N}{2}\xi\lambda_1^+}~,
\eeq
	while the mass of other two states is given by
\beq
\label{mUm}
	\mUm ~~=~~ g_1 \sqrt{\frac{N}{2}\xi\lambda_1^-}~,
\eeq
	where $ \lambda_1^\pm $ are two roots of the quadratic equation
\beq
	\lambda_i^2  -  \lambda_i(2 + \omega^2_i)  +   1  =  0\,,
\label{queq}
\eeq
	for $i =1$, where we introduced two \ntwo supersymmetry breaking parameters associated
	with the U(1) and SU($N$) gauge groups, respectively,
\beq
	\omega_1  =  \frac{g_1\mu_1}{\sqrt{\xi}}\,,\qquad
	\omega_2  =  \frac{g_2\mu_2}{\sqrt{\xi}}\,.
\label{omega}
\eeq
	Other $ 2(N^2 - 1) $ states acquire mass
\beq
\label{mNp}
	\mNp ~~=~~ g_2 \sqrt{\xi\lambda_2^+} ~,
\eeq
	while the remaining $ 2(N^2 - 1) $ states become massive, with mass
\beq
\label{mNm}
	\mNm ~~=~~ g_2 \sqrt{\xi\lambda_2^-} ~,
\eeq
	where $ \lambda_2^\pm $ are two roots of the quadratic equation \eqref{queq} for $ i = 2 $.
	Note that all states come either as singlets or adjoints with respect to the unbroken
	\cfl.

	When the SUSY breaking parameters $ \omega_i $ vanish, the masses $ \mUp $ and $ \mUm $
	coincide with the U(1) gauge boson mass \eqref{MU}.
	The corresponding states form a bosonic part of a long \ntwo massive U(1) vector supermultiplet
	\cite{VY}.	

	If $\omega_1 \neq 0 $ this supermultiplet splits into a \none vector multiplet, with mass $ \MU $,
	and two chiral multiplets, with masses $ \mUp $ and $ \mUm $. 
	The same happens with the states with masses $ \mNp $ and $ \mNm $, see Eqs.~\eqref{mNp} and  \eqref{mNm}.
	With vanishing $ \omega $'s they combine into the bosonic parts of $ (N^2 - 1) $ \ntwo vector supermultiplets
	with mass $ \MN $.
	If $ \omega_i \neq 0 $ these multiplets split into $ (N^2 - 1) $ \none vector multiplets (for the 
	SU($N$) group) with mass \eqref{MN} and $ 2(N^2 - 1) $ chiral multiplets with masses 
	$ \mNp $ and $ \mNm $.
	Note that the same splitting pattern was found in \cite{VY} in the Abelian case.

	Let us take a closer look at the spectrum obtained above in the limit of large \ntwo supersymmetry 
	breaking parameters $\omega_i$, $\omega_i \gg 1 $.
	In this limit the larger masses $ \mUp $ and $\mNp$ become
\beq
\label{amass}
	\mUp ~=~ \MU \omega_1 ~=~ g_1^2\sqrt{\frac{N}{2}}\mu_1 ~,
	\qquad
	\mNp ~=~ \MN \omega_2 ~=~ g_2^2 \mu_2~.
\eeq
	In the limit $\mu_i \to \infty$ these are the masses of the heavy adjoint scalars $ \aU $ and
	$ \aN $.
	At $ \omega_i \gg 1 $ these fields decouple and can be integrated out.

	The low-energy theory in this limit contains massive gauge \none multiplets and chiral multiplets
	with the lower masses $ m^- $. 
	Equation \eqref{queq} gives for these masses
\beq
\label{light}
	\mUm ~=~ \frac{\MU}{\omega_1} ~=~ \sqrt{\frac{N}{2}}\, \frac{\xi}{\mu_1}~,
	\qquad
	\mNm ~=~ \frac{\MN}{\omega_2} ~=~ \frac{\xi}{\mu_2}~.
\eeq
	In particular, in the limit of infinite $ \mu_i $ these masses tend to zero. 
	This reflects the presence of the Higgs branch in \none SQCD.

	The Higgs branch poses a problem for the $ \mu \to \infty $ limit \cite{SYnone},
	due to the presence of massless quark states.
	These states obscure world-sheet physics of the non-Abelian strings. 
	In particular, the strings become infinitely thick, and higher-derivative corrections
	of the effective theory become important.
	The maximal critical value of those values of $ \mu $ where the world sheet theory can be 
	trusted was found in \cite{SYnone}
\[
	\mu_2^* ~~=~~ \frac{\xi^{3/2}}{\left(\Lambda_\text{SU($N$)}^{\mc{N}=1}\right)^2}~,
\]
	where $ \Lambda_\text{SU($N$)}^{\mc{N}=1} $ is the scale of \none SQCD to which the theory
	\eqref{theory} flows in the $ \mu \to \infty $ limit
\[
	\left(\Lambda_\text{SU($N$)}^{\mc{N}=1}\right)^{2N} ~~=~~ \mu_2^N\, \LN^N~.
\]

%
%
\section{Vortex solutions in the bulk}
\label{vortex}
\setcounter{equation}{0}

	The theory with \ntwo supersymmetry admits non-Abelian string solutions in the bulk
\cite{HT1,ABEKY,SYmon,HT2}.
	The presence of the U(1) gauge factor allows for non-trivial winding of the solution
	at infinity.
	These non-Abelian vortices however are different from the conventional ANO strings \cite{ANO}, 
	since they involve winding in both factors SU($N$) and U(1) of the broken gauge group
\[
	\varphi_{\rm string} ~~=~~ \sqrt{\xi}\,{\rm diag}(1, 1, \dots, e^{i\alpha}), 
		\qquad\qquad \text{at $x \to \infty$}
\]
	where $ \alpha $ is the angle in the plane orthogonal to the string.
	These are the so-called $ Z_N $ strings.

	The fact that in theory \eqref{theory} supersymmetry is broken down to \none does not prevent
	one from having BPS strings, provided that one of the quark masses 
	coincide with the critical point of the superpotential \cite{Edalati}.
	Obviously, this is fulfilled with our choice of the quark masses \eqref{qmasses} and superpotential
	\eqref{none}.
	The bosonic part of the classical string solution is then the same as in \ntwon, and hence can
	be used ``as is'';
	 see  \cite{SYrev} and \cite{SYhet} where this solution
is found in \none bulk theory for the case of the U(2) gauge group.

	We take the {\it ansatz} where half of the quark fields vanish, together with the adjoint matter
\begin{align*}
	q       & ~~=~~ \ov{q} ~~\equiv~~ \varphi~,  \\[2mm]
	\wt{q}  & ~~=~~ \ov{\wt{q}} ~~=~~ 0~, \\[2mm]
	\aU     & ~~=~~ \aN ~~=~~ 0~. 
\end{align*}
	With this {\it ansatz}, the bosonic action \eqref{theory} takes a particularly simple form
\begin{align}
\notag
	S ~~=~~ & \int d^4x\, 
	\Biggl\{  \frac{1}{4g^2}\left( F_{\mu\nu}^a \right)^2  ~+~ 
		 \frac{1}{g_1^2}\left( F_{\mu\nu}^{\rm U(1)} \right)^2 ~+~
		 \left| \nabla_\mu \varphi^A \right|^2 ~+~ \\[3mm]
\label{redmodel}
	        & \phantom{ \int d^4x\, \Biggl\{ }
		~+~
		 \frac{g_2^2}{2} \left( \ov{\varphi}{}_A\,T^a\varphi^A \right)^2 ~+~
		 \frac{g_1^2}{8} \left( \ov{\varphi}{}_A\, \varphi^A ~-~ N\,\xi \right)^2
	\Biggr\} \,.
\end{align}
	The profile solution for the string can be written as \cite{ABEKY}
\begin{align}
\notag
	\varphi   ~~=~~  &
		\lgr \begin{matrix}
			\phi_2(r) & 0     & \dots      & 0      \\
			\dots     & \dots & \dots      & \dots  \\
			0         & \dots & \phi_2(r)  & 0      \\
			0         &  0    & \dots      & e^{i\alpha}\phi_1(r) 
		     \end{matrix} \rgr
	\\
\label{string_reg}
	\\[-0.7cm]
\notag
	A_i^\text{SU($N$)}  ~~=~~
		\frac{1}{N} & \lgr \begin{matrix}
        			    	1       &   \dots   &  0       &   0   \\
        				\dots   &   \dots   &  \dots   & \dots \\
        				0       &   \dots   &  1       &   0   \\
        				0       &     0     &  \dots   & - (N-1) 
	   		         \end{matrix} \rgr
		(\p_i\alpha)\bigl( -1 ~+~ f_{N}(r) \bigr)
	\\
\notag
	A_i^{\rm U(1)}  ~~=~~ & \frac{1}{N}(\p_i\alpha)\lgr 1 ~-~ f(r) \rgr \cdot \mathlarger{\mathbf{1}}~,
	\qquad 
	A_0^{\rm U(1)} ~=~ A_0^\text{SU($N$)} ~=~ 0~.
\end{align}
	Here $ i $ labels the coordinates in the orthogonal plane, and $ r $ and $ \alpha $
	are the polar coordinates in this plane.
	The functions $ \phi_1(r) $ and $ \phi_2(r) $ determine the profiles of the scalar quarks
	in the orthogonal plane, while $ f(r) $ and $ f_{NA}(r) $ are the profiles of the gauge
	fields. 
	This {\it ansatz} describes strings with the winding number $ k = 1 $.
	The profiles satisfy the first-order differential equations, which follow from the BPS
	equations upon substitution of the above {\it ansatz} \cite{MY,ABEKY}:
\begin{align}
\notag
&	\p_r\, \phi_1(r) ~-~ \frac{1}{Nr}\, \lgr f(r) ~+~ (N-1)f_{N}(r) \rgr \phi_1(r) ~~=~~ 0, \\
\notag
&	\p_r\, \phi_2(r) ~-~ \frac{1}{Nr}\, \lgr f(r) ~-~ f_{N}(r) \rgr \phi_2(r) ~~=~~ 0 ,\\
\label{foes}
&	\p_r\, f(r) ~-~ r\, \frac{N g_1^2}{4} \lgr (N-1)\phi_2(r)^2 ~+~ \phi_1(r)^2 ~-~ N\xi \rgr ~~=~~ 0 , \\
\notag
&	\p_r\, f_{N}(r)  ~-~  r\, \frac{g_2^2}{2} \lgr \phi_1(r)^2 ~-~ \phi_2(r)^2 \rgr ~~=~~ 0~,
\end{align}
	with the boundary conditions
\begin{align}
\label{boundary}
	\phi_1(0) & ~~=~~  0\text,                   & \phi_2(0) & ~~\neq~~ 0\text,  &
	\phi_1(\infty) & ~~=~~ \sqrt{\xi} \text,     & \phi_2(\infty) & ~~=~~ \sqrt{\xi}\text, \\
\notag
	f_{N}(0) & ~~=~~ 1\text,                   & f(0) & ~~=~~ 1\text,   &
	f_{N}(\infty) & ~~=~~ 0 \text,            &  f(\infty) & ~~=~~ 0\text.
\end{align}
	Under the latter conditions, the quark profile $ \phi_1(r) $ is required to vanish at the origin, while 
	$ \phi_2(r) $ is not restricted at $ r = 0 $, and, generally speaking, does not vanish.
	The $ Z_N $ strings have the tension 
\[
	T_1  ~~=~~ 2\pi\xi~,
\]
	while for the ANO strings one has
\[
	T_{\rm ANO} ~~=~~ 2\pi N \xi~.
\]
	In this sense, the strings \eqref{string_reg} can be viewed as elementary.

	The solution \eqref{string_reg} breaks \cfl down to SU($N-1$)$\times$U(1). 
	This means that the string acquires orientational moduli living in 
\beq
\label{modulispace}
	\frac{\text{SU($N$)}}
            {\text{SU($N-1$)} \times {\rm U(1)}}         ~~\sim~~  \text{CP($N-1$)}~
\eeq
	and becomes {\it bona fide} non-Abelian.

	The orientational degrees of freedom can be defined as follows. 
	Since the solution \eqref{string_reg} is one of a family of string solutions, 
	{\it i.e.} a representative, one can recover the entire family
	by acting 
	with the diagonal color-flavor rotations preserving 
	the vacuum \eqref{qVEV}.
	For convenience we pass hereforth to the singular gauge where the scalar fields do not wind, but 
	the gauge fields have winding around the origin; in this gauge the family
	of solutions takes the form
\begin{align}
\notag
	\varphi ~~=~~ & U\, \lgr \begin{matrix}
			   	\phi_2(r)  & 0  & \ldots & 0 \\
				\ldots  &  \ldots & \ldots & \ldots \\
				0  & \ldots      & \phi_2(r) &  0 \\
				0  & 0           & \ldots  &  \phi_1(r) 
			   \end{matrix}        \rgr     
			U^{-1} \,,\\[2mm]
\label{nastr}
	A_i^\text{SU($N$)} ~~=~~ \frac{1}{N}\, &\, U\, \lgr \begin{matrix}
					          	1  & \ldots & 0 & 0 \\
						  	\ldots & \ldots & \ldots & \ldots \\
							0  & \ldots  & 1  &  0 \\
							0  & 0   & \ldots  &  - (N-1) 
					         \end{matrix} \rgr  \, U^{-1} (\p_i \alpha)\, f_{N}(r)\,,  \\[2mm]
	A_i^{\rm U(1)} ~~=~~ -\, & \frac{1}{N}\, (\p_i \alpha)\, f(r) \cdot \mathlarger{\mathbf{1}}~, 
	\qquad\qquad
			A_0^{\rm U(1)} ~~=~~ A_0^\text{SU($N$)} ~~=~~ 0~.
\end{align}
	Here $ U $ is a unitary color-flavor rotation matrix from \cfl\hspace{-1ex}.	
	Since a string solution breaks \cfl\hspace{-1ex}, the effective two-dimensional theory on the
	string is described by a CP($N-1$) theory of orientational moduli, 
	see Eq.~\eqref{modulispace}.
	It is convenient to describe this theory in terms of more explicit orientational variables $n^l$, which are
	related to the rotation matrix $ U $ in \eqref{nastr} as follows
\beq
\label{n}
	\frac{1}{N}\, U \, \lgr \begin{matrix}
				  1  & \ldots & 0 & 0 \\
				  \ldots & \ldots & \ldots & \ldots \\
				  0 & \ldots & 1 & 0  \\
				  0 & 0 & \ldots & -(N-1) 
				\end{matrix} \rgr
			U^{-1}  
	~~=~~
	-\, n^i\,\ov{n}{}_l  ~~+~~ \frac{1}{N}\cdot{\mathlarger{\mathbf{1}}}{}^i_{~l} ~,
\eeq
	where matrix notation is used on the left-hand side. 
	These variables are subject to the CP($N-1$) conditions, which can be converted into
	the constraint
\[
	\ov{n}{}_l \cdot n^l ~~=~~ 1 \text.
\]
	In addition, there is a freedom of multiplication by one common complex phase
	(for example, one of $n^l$ can be chosen real). 
	In the gauged formulation of the sigma model this latter phase ambiguity 
	arises as a freedom of gauge.
	The number of degrees of freedom is therefore $ 2(N-1) $.

	Using this parametrization of CP($N-1$), one rewrites the string solution
	\eqref{nastr} as
\begin{align}
\notag
	\varphi & ~~=~~ 
		\phi_2 ~+~ n\nbar\, \bigl( \phi_1 ~-~ \phi_2 \bigr) 
	\\[2mm]
\notag
		& 
		 ~~=~~  
			\frac{1}{N}\bigl( \phi_1 ~+~ (N-1)\phi_2 \bigr)
  			       ~+~ \bigl( \phi_1 ~-~ \phi_2 \bigr)
			           \lgr n\nbar ~-~ 1/N \rgr \,,
	\\[2mm]
\label{str}
	A_i^\text{SU($N$)} & ~~=~~ \varepsilon_{ij}\, \frac{x^j}{r^2}\, f_{N}(r)
				\lgr n\nbar ~-~ 1/N \rgr\,,
	\\[2mm]
\notag
	A_i^{\rm U(1)} & ~~=~~ \frac{1}{N}\varepsilon_{ij}\, \frac{x^j}{r^2}\, f(r)~,
\end{align}
	where we use matrix notation on both sides of the equations. 
	
	To derive the effective theory on the string, one assumes that 
	the orientational moduli $ n^l $ are slowly varying functions of the world-sheet coordinates
	$ x_k $, $ k = 0, 3 $. 
	The former then become fields living on the world sheet. 
	Since they parameterize the string zero modes, there is no potential
	term in this sigma model.

	To obtain the kinetic term for the world sheet theory, one substitutes the string {\it ansatz}
	into action \eqref{redmodel}, with moduli $ n^l $ now adiabatically
	depending on $ x_0 $, $ x_3 $.
	However, in the presence of the latter dependence, the solution \eqref{str} has to be altered.
	The reason is seen from Eq.~\eqref{nastr}, where the transformation parameter 
	$ U $ is no longer global, but rather is a function of the world sheet coordinates.
	This leads to the fact that the SU($N$) gauge field, in addition to the transversal components
	given in \eqref{str}, acquires longitudinal components as well.
	The {\it ansatz} for the longitudinal components of the gauge field should
	therefore include the derivatives of $ n^l $ and can be chosen as 
\[
	A_\mu^\text{SU($N$)} ~~=~~ i\, \left[\, n\nbar, \p_\mu(n\nbar)\, \right] \rho(r)~,   \qquad\qquad\qquad \mu~=~0, 3~,
\]
	where $ \rho(r) $ is the corresponding profile function, which can be determined by
	a minimization procedure \cite{SYrev}
\[
	\rho(r) ~~=~~ 1 ~-~ \frac{\phi_1}{\phi_2}~.
\]
	Substituting the above {\it ans\"{a}tze} into bosonic action \eqref{redmodel} one arrives
	at
\begin{align}
\label{cp}
	S_{\rm bos}^{1+1} ~~=~~ 2\beta\, \int\, dt\,dz 
					\Bigl\{\, \left|\p n^l\right|^2    
						  ~+~  \left(\nbar \p_k n\right)^2\,
					\Bigr\}~,
\end{align}
	where the coupling constant $ \beta $ is 
\beq
\label{beta}
	\beta ~~=~~ \frac{2\pi}{g_2^2}\,.
\eeq
	This relation establishes a classical connection for the two- and four-dimensional
	 coupling constants.
	Quantum mechanically, both of the couplings run, in particular, $ \beta $ is asymptotically
	free \cite{P75}.
	The scale at which equality \eqref{beta} holds is determined by the ultraviolet cut-off
	of the effective theory, which is given by the inverse thickness of the string $ g_2 \sqrt{\xi} $.
	The running of the coupling $ \beta $ is 
\beq
\label{asyfree}
	4 \pi \beta ~~=~~ N\,  \ln \frac{E}{\Lambda_\sigma}~,
\eeq
	where $ \Lambda_\sigma $ is the dynamical scale of the sigma model, 
\beq
\label{lambdasig}
	\Lambda_\sigma ~~=~~ g_2^N\, \xi^{\frac{N}{2}}\, e^{-\frac{8\pi^2}{g_2^2}} ~~=~~ \LN~.
\eeq

%
%
\section{Zero modes of the unbroken \boldmath\ntwo theory}
\label{zeromodes}
\setcounter{equation}{0}

	In \ntwo theory, the lightest modes are the zero modes. 
	The effective action on the string world sheet can be built by calculating the zero mode overlap 
	by substituting them into the theory \eqref{theory}, in a way similar to deriving \eqref{cp}.
	Now, however, we are more interested in the fermionic zero modes, since it is the latter that
	are modified in the presence of  the \ntwon-breaking superpotential \eqref{none}.

	Our string solution  is 1/2-BPS, that is, half of the supercharges vanish when acting on the solution.
	The other half does not vanish and generate the supertranslational zero modes, the superpartners 
	to the translational modes.
	In fact, the latter two are related to each other in a simple manner \cite{Edalati},
\beq
\label{bos_ferm_rel}
	\psi_{\text{s-trans}} ~~=~~  \delta A \cdot \zeta~,
\eeq
	where $ \delta A $ symbolically denotes the corresponding bosonic mode.
	Here $ \zeta $ is the 
	fermionic superpartner to the translational moduli $ x_0^i $, $ i = 1, 2 $, representing the position of the 
	center of the string.
	Similar relation holds between the orientational and superorientational modes.

	The fermionic supertranslational zero modes 
for the U(1)$ \times $ SU(2) non-Abelian string were found in \cite{SYhet}.
	Now we will generalize this to the case of SU($N$)$\times$U(1).
	The fermionic part of the theory \eqref{theory} is
\begin{align}
\notag
\mc{L}_{\rm 4d} & ~~=~~ \frac{2i}{g_2^2} \Tr\, \ov{\lN_f \slashed{\md}} \lambda^{f\text{SU($N$)}}
		~+~ \frac{4i}{g_1^2} \ov{\lU_f \slashed{\p}} \lambda^{f{\rm U(1)}}
		~+~ \Tr\, i\, \ov{\psi \slashed{\md}} \psi  
		~+~ \Tr\, i\, \wt{\psi} \slashed{\md} \ov{\wt{\psi}}
		\\[3mm]
\notag
		& 
		~+~
		i\sqrt{2}\, \Tr \lgr \ov{q}{}_f \lambda^{f{\rm U(1)}}\psi 
				  ~+~ \wt{\psi} \lU_f q^f  
				  ~+~ \ov{\psi \lU_f} q^f
				  ~+~ \ov{q^f \lU_f \wt{\psi}} 
				\rgr
		\\[3mm]
\label{fermact}
		&
		~+~
		i\sqrt{2}\, \Tr \lgr \ov{q}{}_f \lambda^{f\text{SU($N$)}} \psi 
					~+~ \wt{\psi} \lN_f q^f
					~+~ \ov{\psi \lN_f} q^f
					~+~ \ov{q^f \lN_f \wt{\psi}}
				\rgr
		\\[3mm]
\notag
		&
		~+~
		i\sqrt{2}\, \Tr\; \wt{\psi} \left( \aU ~+~ \aN \right) \psi  
		~+~ 
		i\sqrt{2}\, \Tr\; \ov{\psi} \left( \baU ~+~ \baN \right) \ov{\wt{\psi}}
		\\[3mm]
\notag
		&
		~-~
		2 \sqrt{\frac{N}{2}} \mu_1 \lgr \left( \lambda^{2\,{\rm U(1)}} \right)^2 
  					    ~+~ \left( \ov{\lambda}{}^{\rm U(1)}_2 \right)^2 \rgr
		~-~
		\mu_2 \Tr \lgr \left( \lambda^{2\,\text{SU($N$)}} \right)^2
			   ~+~ \left( \ov{\lambda}{}^\text{SU($N$)}_2 \right)^2 \rgr\,.
\end{align}
	The long bars in this equation indicate that each corresponding variable comes with a bar. 
	In the case of the derivatives $\overline{\slashed{\partial}}$ the bar signifies that 
	the derivative is contracted with $\overline{\sigma}{}^\mu_{\dot{\alpha}\alpha} $ rather 
	than $\sigma_\mu^{\alpha\dot{\alpha}}$.
	We use the matrix color-flavor notation for fermions; the traces run
	over the corresponding color-flavor indices.
	The squark fields are written as SU(2)$_R$ doublets of the \ntwo theory,
	$ q^f ~=~ (q, \ov{\wt{q}}) $. 
	The index $ f = 1, 2 $ labels two supersymmetries which are present in the \ntwo limit. 
	In particular, $ f = 2 $ gauginos, which are part of the adjoint multiplets, are given mass terms in \eqref{fermact}.
	See Appendix  for other notations.
	
	In order to find the zero modes, one generally speaking has to solve the Dirac equations.
	However, in the presence of supersymmetry,  one can simply apply the supersymmetry transformations
	to the bosonic solution to obtain the fermionic zero modes.
	This follows from the fact that the fermionic and bosonic modes are related to each other, 
	see Eq.~\eqref{bos_ferm_rel}.
	In the bulk, the supersymmetry transformations are
\begin{align}
\notag
	\delta\lambda^{f\alpha}_{\rm U(1)} & ~~=~~ \frac{1}{2} \left(\sigma_\mu \ov{\sigma}{}_\nu \epsilon^f \right)^\alpha 
							        F_{\mu\nu}^{\rm U(1)}  
						~+~ \epsilon^{\alpha p} D^{{\rm U(1)}\,m} \left( \tau^m \right)^f_{\ p}
						~+~ \dots~,
\\
\notag
	\delta\lambda^{f\alpha}_\text{SU($N$)} & ~~=~~ \frac{1}{2} \left(\sigma_\mu \ov{\sigma}{}_\nu \epsilon^f \right)^\alpha
								F_{\mu\nu}^\text{SU($N$)}
						~+~ \epsilon^{\alpha p} D^{\text{SU($N$)}\,m} \left( \tau^m \right)^f_{\ p}
						~+~ \dots~,
\\[-0.7cm]
\label{transf}
\\
\notag
	\delta\ov{\wt{\psi}}{}_{\dot{\alpha}}^{kA} & ~~=~~ i\sqrt{2}\, 
				\left( \ov{\slashed{\nabla}}_{\dot{\alpha}\alpha} q_f \right)^{kA} \epsilon^{\alpha f} 
						~+~ \dots~,
\\
\notag
	\delta\ov{\psi}{}_{\dot{\alpha} A k} & ~~=~~ i\sqrt{2}\, 
				\left( \ov{\slashed{\nabla}}_{\dot{\alpha}\alpha} \ov{q}{}_f \right)_{Ak} \epsilon^{\alpha f}
						~+~ \dots ~.
\end{align}
	The parameter of supertransformations is $ \epsilon^{\alpha f} $. 
	The ellipses stand for the adjoint scalar contributions, which vanish on our string solution. 
	We have used the matrix notation in \eqref{transf} for both the SU($N$) gauginos and quark fields, where for the latter
	we present indices explicitly.
	The $ D $-terms in Eq.~\eqref{transf} are
\begin{align}
\notag
	& D^{{\rm U(1)}\,1}  ~+~  i D^{{\rm U(1)}\,2} ~~=~~ 0~,      
	&& D^{{\rm U(1)}\,3} ~~=~~ -i\, \frac{g_1^2}{4}\, \lgr \Tr |\varphi|^2 ~-~ N\,\xi \rgr,
	\\[3mm]
\label{dterm}
	& D^{\text{SU($N$)}\,1}  ~+~  i D^{\text{SU($N$)}\,2} ~~=~~ 0~,    
	&& D^{\text{SU($N$)}\,3} ~~=~~ -i\, g_2^2\; \Tr \lgr \ov{\varphi}\, T^a \varphi \rgr T^a~.
\end{align}
	The supertransformations generated using parameters $ \epsilon^{12} $ and $ \epsilon^{21} $ act trivially on the
	BPS string in theory with the Fayet--Iliopoulos $ D $-term \cite{VY, SYhet}.
	The other two supertransformations which are associated with the parameters $ \epsilon^{11} $ and $ \epsilon^{22} $,
	generate supertranslational zero modes, and in fact these parameters are identified with the corresponding world sheet
	coordinates
\beq
\label{zeta}
	\zl ~~=~~ \epsilon^{11}~,  \qquad\qquad   \zr ~~=~~ \epsilon^{22}~.
\eeq
	The zero modes are obtained straightforwardly by substituting the bosonic string solution into Eq.~\eqref{transf}
\begin{align}
\label{N2_strans}
\notag
\ov{\psi}_{\dot{2}}	& ~~=~~  -\,  2\sqrt{2}\, \frac{x_1 ~+~ i x_2}{N r^2} \,
		\lgr \frac{1}{N} \phi_1 ( f + (N-1) f_N ) ~+~ \frac{N-1}{N} \phi_2 ( f - f_N )  \right.
		\\[2mm]
\notag
			& \phantom{~~=~~  -\,  2\sqrt{2}}
			~+~ \left( n\nbar ~-~ 1/N \right )
			\Bigl\{ \phi_1 ( f + ( N-1 ) f_N ) ~-~ \phi_2 ( f - f_N) \Bigr\}
		\left. \rgr\, \zeta_L \,,
		\\[2mm]
\notag
\ov{\wt{\psi}}_{\dot{1}} & ~~=~~    2\sqrt{2} \, \frac{x_1 ~-~ i x_2}{N r^2} \,
		\lgr \frac{1}{N} \phi_1 ( f + (N-1) f_N ) ~+~ \frac{N-1}{N} \phi_2 ( f - f_N ) \right.
		\\[2mm]
\notag
			& \phantom{~~=~~  -\,  2\sqrt{2}}
			~+~ \left( n\nbar ~-~ 1/N \right )
			\Bigl\{ \phi_1 ( f + ( N-1 ) f_N ) ~-~ \phi_2 ( f - f_N) \Bigr\}
		\left. \rgr\, \zeta_R\,,
		\\[2mm]
\lambda^{11\ \rm U(1)} 	& ~~=~~ -\, \frac{i g_1^2}{2} \lgr (N-1)\phi_2^2  ~+~ \phi_1^2 ~-~ N\xi \rgr \, \zeta_L \,,
		\\[2mm]
\notag
\lambda^{22\ \rm U(1)} 	& ~~=~~ +\, \frac{i g_1^2}{2} \lgr (N-1)\phi_2^2  ~+~ \phi_1^2 ~-~ N\xi \rgr \, \zeta_R \,,
		\\[2mm]
\notag
\lambda^{11\ \text{SU($N$)}}	& ~~=~~ -\, {i g_2^2}\, ( n\nbar ~-~ 1/N )\, \lgr \phi_1^2 ~-~ \phi_2^2 \rgr\, \zeta_L\,,
		\\[2mm]
\notag
\lambda^{22\ \text{SU($N$)}}	& ~~=~~ +\, {i g_2^2}\, ( n\nbar ~-~ 1/N )\, \lgr \phi_1^2 ~-~ \phi_2^2 \rgr\, \zeta_R
	~,
\end{align}
	where we list only the non-vanishing components of the fermions.
	
	The superorientational zero modes are obtained using the supertransformations generated by 
	$ \epsilon^{21} $ and $ \epsilon^{12} $. This method was suggested in 
\cite{SYmon} and was used to determine fermionic superorientational zero modes
in the \ntwo theory with the U(2) gauge group. Here we develop a U$(N)$ generalization of results
in \cite{SYmon}.
As was already mentioned, a direct substitution of the bosonic solution \eqref{str} into the transformations
	\eqref{transf} with these parameters would produce a vanishing result.
	The zero modes are in fact proportional to the $ x^0 $, $ x^3 $-derivatives of the orientational moduli $ n^l $.
	If one assumes a slow longitudinal dependence of the orientational coordinates in \eqref{str}, then Eq.~\eqref{transf}
	yields
\begin{align*}
	\ov{\psi}{}_{\dot{\alpha}Ak} & ~~=~~  \phantom{-\, } i\sqrt{2}\, \delta_{\dot{\alpha}}^{\ \dot{2}}\;
					\frac{\phi_1^2 ~-~ \phi_2^2}{\phi_2}\cdot n\nbar\, \p_L (n\nbar)\cdot \epsilon^{21}~,
	\\[2mm]
	\ov{\wt{\psi}}{}_{\dot{\alpha}}^{kA} & ~~=~~ -\,  i\sqrt{2}\, \delta_{\dot{\alpha}}^{\ \dot{1}}\;
					\frac{\phi_1^2 ~-~ \phi_2^2}{\phi_2}\cdot \p_R (n\nbar)\, n\nbar\cdot \epsilon^{12}~,
	\\[2mm]
	\lambda^{f\alpha\, \text{SU($N$)}} & ~~=~~ 
		2\, \frac{\phi_1}{\phi_2}\, f_N
	\lgr \begin{matrix}
			-\, \frac{\displaystyle x^1 - i\, x^2}{\displaystyle r^2}\cdot
				n\nbar\, \p_L(n\nbar)\cdot \epsilon^{21}                     &  0  \\[2mm]
			0 &
			    \frac{\displaystyle x^1 + i\, x^2}{\displaystyle r^2}\cdot 
				\p_R (n\nbar)\, n\nbar\cdot \epsilon^{12}
	     \end{matrix} \rgr^{f\alpha},
\end{align*}	
	where
\[	
	\p_R ~~=~~ \p_0 ~+~ i\, \p_3 ~, \qquad\qquad  \p_L ~~=~~ \p_0 ~-~ i\, \p_3~.
\]
	Using the world sheet supersymmetry transformations, see {\it e.g.} Eq.~\eqref{susy_ntwot}, one finds that 
	the derivatives of the orientational moduli generate the fermionic superpartners $ \xi^l $ of the latter
\begin{align*}
	i \sqrt{2}\; \p_R (n \nbar)\, n\nbar \cdot \epsilon^{12} & 
		~~=~~ \xi_R \nbar~,  \\[2mm]
	i \sqrt{2}\; n\nbar\, \p_L (n\nbar) \cdot \epsilon^{21} & 
		~~=~~ n \bxil~.
\end{align*}
	The variables $ \xi_{R,L} $ are constrained to be orthogonal to the orientational moduli $ n^l $, which is the
	supersymmetric generalization of the CP($N-1$) condition $ |n|^2 = 1 $:
\beq
\label{constr}
	\nbar{}_l\, n^l ~~=~~ 1\,, \qquad\qquad    \nbar_l\, \xi^l  ~~=~~ \ov{\xi}{}_l\, n^l  ~~=~~ 0~.
\eeq
	One then arrives at the following result for the superorientational modes:
\begin{align}
\label{N2_sorient}
\notag
\overline{\psi}_{\dot{2}Ak} & ~~=~~ \frac{\phi_1^2 ~-~ \phi_2^2}{\phi_2} \cdot n \overline{\xi}_L  \,,
 \\[2mm]
\notag
\overline{\wt{\psi}}_{\dot{1}}^{kA}  & ~~=~~ - \frac{\phi_1^2 ~-~ \phi_2^2}{\phi_2} \cdot \xi_R \nbar \,, \\[2mm]
\lambda^{11\ \text{SU($N$)}} & ~~=~~ i \sqrt{2}\, \frac{ x^1 ~-~ i\, x^2 }{r^2} 
						  \frac{\phi_1}{\phi_2} f_N \cdot n \overline{\xi}_L \,,\\[2mm]
\notag
\lambda^{22\ \text{SU($N$)}} & ~~=~~ - i \sqrt{2}\, \frac{ x^1 ~+~ i\, x^2 }{r^2} 
						    \frac{\phi_1}{\phi_2} f_N \cdot \xi_R \nbar 
	~,
\end{align}
	where, again, only non-zero components are shown. 
	The results \eqref{N2_strans} and \eqref{N2_sorient} qualitatively agree with those in Ref.~\cite{Edalati},
	where, however, only the case of equal coupling constants $ g_1 = g_2 $ was considered.
	In that case one only needs a single quark profile function rather than two $ \phi_1 $ and $ \phi_2 $.
	In general, even if the \sunu theory is obtained from a broken SU$ (N+1) $, 
	one is not free to assume the coupling constants to be equal, due to their running above $ \sqrt{\xi} $.

	From the action \eqref{fermact} one finds the Dirac equations which the zero modes should satisfy,
\begin{align}
\notag
	&\phantom{-i}
	\frac{4i}{g_1^2}\, \left( \ov{\slashed{\p}}\lambda^{f {\rm U(1)}} \right) 
		~+~  i \sqrt{2}\, \Tr\lgr \ov{\psi} q^f  ~+~ \ov{q}{}^f \ov{\wt{\psi}} \rgr
		~-~ 4\, \delta_2^{\ f}\sqrt{\frac{N}{2}} \mu_1\, \ov{\lambda}^{U(1)}_2  ~~=~~ 0\,, 
		\\[2mm]
\notag
	&\phantom{-i}
	\frac{i}{g_2^2}\, \left( \ov{\slashed{\md}}\lambda^{f \text{SU($N$)}}\right)^a 
		~+~ i \sqrt{2}\, \Tr\lgr \ov{\psi}T^a q^f  ~+~  \ov{q}{}^f T^a \ov{\wt{\psi}} \rgr
		~-~ \delta_2^{\ f} \mu_2\, \ov{\lambda}{}_2^{a \text{SU($N$)}}  ~~=~~ 0 \,,\\[2mm]
\notag
	&
	-i\, \ov{\psi} \overleftarrow{\ov{\slashed{\nabla}}}
		~+~ i \sqrt{2} \lgr \ov{q}{}_f \left\{ \lambda^{f {\rm U(1)}} + \lambda^{f \text{SU($N$)}} \right\}
					~+~ \wt{\psi} \left\{ \baU  ~+~ \baN \right\} \rgr    ~~=~~ 0\,, \\[2mm]
\label{dirac}
	&\phantom{-i}
	i\, \slashed{\nabla} \ov{\wt{\psi}} 
		~+~ i \sqrt{2} \lgr \left\{ \lambda_f^{\rm U(1)} ~+~ \lambda_f^\text{SU($N$)} \right\} q^f
					~+~ \left\{ \baU ~+~ \baN \right\}\psi \rgr  ~~=~~ 0\,, \\[2mm]
\notag
	&\phantom{-i}
	i \ov{\slashed{\nabla}}\psi 
		~+~ i \sqrt{2} \lgr \left\{ \ov{\lambda}{}^{\rm U(1)}_f ~+~ \ov{\lambda}{}^\text{SU($N$)}_f \right\} q^f
					~+~ \left\{ \aU ~+~ \aN \right\} \ov{\wt{\psi}} \rgr   ~~=~~ 0 \,,\\[2mm]
\notag
	&
	-i\, \wt{\psi} \overleftarrow{\slashed{\nabla}}
		~+~ i \sqrt{2} \lgr \ov{q}{}^f \left\{ \ov{\lambda}{}_f^{\rm U(1)} ~+~ \ov{\lambda}{}_f^\text{SU($N$)} \right\}
					~+~ \ov{\psi} \left\{ \aU ~+~ \aN \right\} \rgr  ~~=~~ 0\,.
\end{align}
	Indeed, it can be readily verified, that the zero modes \eqref{N2_strans} and \eqref{N2_sorient} satisfy these
	Dirac equations in the limit when the supersymmetry breaking parameters $ \mu_1 $ and $ \mu_2 $ vanish.

	In the remainder of this section we present the effective \ntwot world sheet theory on the string.
	This is done by substituting the above zero modes into the kinetic terms of our theory \eqref{theory}.
	Before delving in this, we recall that in Section~\ref{vortex} the presence of the orientational modes
	forced us to include the longitudinal components of the SU($N$) gauge field,
\beq
\label{Aorlong}
	A_\mu^\text{SU($N$)} ~~=~~ i\, \left[\, n\nbar, \p_\mu(n\nbar)\, \right] \rho(r)~,   \qquad\qquad\qquad \mu~=~0, 3~.
\eeq
	Similarly, the dependence of the translational moduli $ x_0^i $ on the world sheet coordinates
	$ x^0 $, $ x^3 $ induces longitudinal contributions in both Abelian and non-Abelian gauge fields,
\begin{align}
\notag
	A_\mu^{\rm U(1)}	& ~~=~~ \epsilon_{ij}\,\frac{ (x^i - x_0^i)\,\p_\mu(x^j - x_0^j)} {r^2}\, f(r)~, \\[2mm]
\label{Atrlong}
	A_\mu^\text{SU($N$)}	& ~~=~~ \epsilon_{ij}\,\frac{ (x^i - x_0^i)\,\p_\mu(x^j - x_0^j)} {r^2}\, f_N(r)~,
				\qquad\qquad \mu~=~0, 3~.
\end{align}

	Now, introducing the string center coordinates $ x_0^1 $ and $ x_0^2 $ into the string solution \eqref{str}, and 
	plugging the latter together with \eqref{Atrlong} and \eqref{N2_strans} into the action \eqref{redmodel}, \eqref{fermact},
	one arrives at the translational sector of the world sheet theory,
\[
	S_{\rm trans} ~~=~~ 2\pi\xi\, \int dt\,dz 
                                       \lgr
					     \frac{1}{2} \left(\p_k \vec{x}_0 \right)^2
					~+~  \frac{1}{2} \ov{\zeta}{}_L\, i\p_R \zeta_L 
					~+~  \frac{1}{2} \ov{\zeta}{}_R\, i\p_L \zeta_R
				       \rgr~,
\]
	after properly normalizing the fermions.

	The non-Abelian nature of the string dynamics is contained in the orientational sector of the world sheet theory.
	Equation~\eqref{cp} shows the bosonic part of this sector. 
	To obtain the fermionic part, one substitutes the superorientational zero modes \eqref{N2_sorient}, together with
	\eqref{Aorlong}, into the kinetic terms of our theory \eqref{redmodel}.
	The result gives the kinetic terms of the fermionic moduli of the CP($N-1$) model, namely,
\[
	2\beta \int d^2x \left\{ \bxil\, i\p_R\, \xil ~+~ \bxir\, i\p_L\, \xir \right\}.
\]
	The remaining parts  of the model are recovered by \ntwot supersymmetry.
	We combine the translational and orientational sectors to obtain
\begin{align}
\notag
\mc{S}_{\rm 1+1}^{\rm (2,2)}  ~~=~~ 
	\int  d^2x
	\Biggl\lgroup\; 
	&
		2\pi\xi \lgr   \frac{1}{2} \left(\p_k \vec{x}_0 \right)^2
				~+~  \frac{1}{2} \ov{\zeta}{}_L\, i\p_R\, \zeta_L 
				~+~  \frac{1}{2} \ov{\zeta}{}_R\, i\p_L\, \zeta_R
			\rgr
	\\[2mm]
\notag
	~~+~~  
	&\;
	2\beta \lgr \left|\p_k n \right|^2  ~+~ \left(\ov{n}\p_k n\right)^2  
		~+~ \ov{\xi}{}_L\, i\p_R\, \xi_L  ~+~ \ov{\xi}{}_R\, i\p_L\,  \xi_R 
		\right . \\[2mm]
\label{str_ntwot}
	&\;
	~~-~
	i \left(\nbar\p_R n\right)\, \bxil\xil ~-~ i \left(\nbar\p_Ln\right) \, \bxir\xir 
	\\[2mm]
\notag
	&\;
	\left .
		~~+~
		\bxil \xir \bxir \xil ~-~ \bxir \xir \bxil \xil
	 \rgr
	\Biggr\rgroup ~.
\end{align}
	Here the contractions of the CP($N-1$) indices are implied in an obvious manner, {\it e.g.}
\[
	(\nbar\, \p_R n)\, \bxil\xil  ~~=~~ (\nbar_l\, \p_R n^l)\, \ov{\xi}{}_{Li}\, \xil^i~.
\]
%
%

	The orientational sector of this theory possesses the following \ntwot supersymmetry
\beqn
\delta n & = & \sqrt{2}\, (\epsilon_R\, \xi_L+\epsilon_L\, \xi_R)\,,
\nonumber\\[3mm]
\delta \nbar & = &  \sqrt{2}\,( \ov{\epsilon_R\, \xi}{}_L+ \ov{\epsilon_L\, \xi}{}_R)\,,
\nonumber\\[3mm]
\delta \xi^l_L & = & i\sqrt{2}\,\bar{\epsilon}_R\nabla_L n^l 
-\sqrt{2}\bar{\epsilon}_L(\bar{\xi}_R\xi_L)n^l\,,
\nonumber\\[3mm]
\delta \bar{\xi}_{Ll} & = & i\sqrt{2}\,\epsilon_R\nabla_L \bar{n}_l 
-\sqrt{2}\epsilon_L(\xi_R\bar{\xi}_L)\bar{n}_l\,,
\nonumber\\[3mm]
\delta \xi^l_R & = & i\sqrt{2}\,\bar{\epsilon}_L\nabla_R n^l 
-\sqrt{2}\bar{\epsilon}_R(\bar{\xi}_L\xi_R)n^l\,,
\nonumber\\[3mm]
\delta \bar{\xi}_{Rl} & = & i\sqrt{2}\,\epsilon_L\nabla_R \bar{n}_l 
-\sqrt{2}\epsilon_R(\xi_L\bar{\xi}_R)\bar{n}_l\,,
\label{susy_ntwot}
\eeqn
where $\epsilon_R\equiv\epsilon_1$ and $\epsilon_L\equiv\epsilon_2$ are parameters of 
two SUSY transformations while $\nabla_k=\p_k-iA_k$ with the gauge potential  given by
\beqn
A_0~+~ iA_3 & = & -\;i\,\bar{n}\p_R n ~-~ \bar{\xi}_R\xi_R,
\nonumber\\[3mm]
A_0~-~ iA_3 & = & -\;i\,\bar{n}\p_L n ~-~ \bar{\xi}_L\xi_L.
\eeqn

%
%
\section{Formulations of \boldmath\ntwoo Theory}
\label{FT}
\setcounter{equation}{0}

	When one breaks the microscopic \ntwo theory down to \nonen, it is natural to expect
that the low-energy theory on the string will be described by an \ntwoo supersymmetric sigma model.
As was argued in Ref.~\cite{Edalati}, and confirmed in \cite{SYhet} for the SU(2) theory,
the world sheet model is actually \CPC rather than just CP($N-1$), by the reason
that the latter does not admit \ntwoo generalizations. 
The extra factor  $C$ comes from a mixing of the orientational and translational degrees of
freedom, which do not interact in the \ntwot theory \eqref{str_ntwot}.

	In the language of two-dimensional superfields, it was discovered in \cite{Edalati} that
the deformation of the \ntwot theory consists of adding a two-dimensional superpotential
\[
	\mathcal{W}_{1+1} ~~=~~ \frac{1}{2}\,\delta\,\Sigma^2~,
\]
(here $ \Sigma $ is an appropriate auxiliary superfield)
and mixing it with the right-handed supertranslational modes $ \zeta_R $. 
This was done in the two-dimensional superfield formulation of \CPCn.
	We will not dwell on details of this formulation, but rather note that the latter
mixing can be presented as
\[
	i\, m_W \ov{\lambda}_L \frac{\p^2 \mc{W}_{1+1}}{\p \sigma^2} \zeta_R  ~+~ \text{H.c.},
\]
	written in our notations. 
	Here $ \lambda_L $ is an auxiliary variable responsible for the CP($N-1$) constraints, as described below.
	In essence, the superfield formulation \emph{is} the gauge formulation, to be discussed
	shortly, with
	all physical and auxiliary fields concisely packed into \ntwoo supermultiplets.
	The \ntwot multiplets split up in pairs of \ntwoo ones, and some physical fields become
	isolated, {\it e.g.} $ \zeta_R $ now lives in a multiplet of its own, not related at all to $ x_0 $.
	This is of course the consequence of the breaking of \ntwot symmetry.

	We start with presenting first the gauge formulation of the \CPC theory,
\begin{align*}
	& S_{1+1}^{(0,2)} ~~=~~ \int d^2 x 
\Biggl \lgroup
	2\pi\xi \lgr \frac 1 2 \left (\p x_0\right)^2 ~+~
		      \frac 1 2 \ov{\zeta}{}_L\, i\p_R\, \zeta_L ~+~
			\frac 1 2 \ov {\zeta}{}_R\, i\p_L\, \zeta_R
		\rgr
	 \\[2mm]
&
	\;\;
	~+~
	2\beta \left| \nabla n \right|^2 ~+~ \frac{1}{4e^2} F_{kl}^2
			~+~ \frac{1}{e^2} \left|\p \sigma\right|^2 ~+~
	4\beta\, |\sigma|^2\, |n|^2 ~+~
	2\, e^2\beta^2 \left( |n|^2 ~-~ 1 \right)^2 \\[2mm]
&
	\;\;
	~+~
	\frac{1}{e^2} \ov{\lambda}{}_R\, i\p_L \lambda_R ~+~
	\frac{1}{e^2} \ov{\lambda}{}_L\, i\p_R \lambda_L ~+~
	2\beta\, \ov{\xi}{}_R\, i\nabla_L\, \xi_R ~+~
	2\beta\, \ov{\xi}{}_L\, i\nabla_R\, \xi_L  \\[2mm]
&
	\;\;
	~+~
	i\,\sqrt{2}\,2\beta \lgr \sigma\, \ov{\xi}{}_R\xi_L ~+~ 
				\ov{\sigma}\,\ov{\xi}{}_L\xi_R \rgr  \\[2mm]
&
	\;\;
	~+~
	i\,\sqrt{2}\,2\beta 
		\lgr \ov{n} \left( \lambda_R \xi_L ~-~ \lambda_L\xi_R \right)
		~-~ \left( \ov{\xi_L\lambda}{}_R ~-~ \ov{\xi_R\lambda}{}_L \right) n \rgr 
	\\[2mm]
&	
	\;\;
	~+~
	2\beta\,
	\lgr 4 \left | \frac{\p \mc{W}_{1+1}}{\p\sigma} \right |^2 
		~+~ i\, m_W \ov{\lambda}_L \frac{\p^2 \mc{W}_{1+1}}{\p \sigma^2} \zeta_R 
	~+~ i\, m_W \ov{\zeta}{}_R \frac{\p^2 \ov{\mc{W}}{}_{1+1}}{\p\ov{\sigma}^2} \lambda_L 
	\rgr
	\Biggr \rgroup~.
\label{cpgf}
\end{align*}
	The sigma model appears in the $ e^2 \to \infty $ limit upon elimination of  auxiliary fields, 
	including the U(1) gauge field.
	The only physical fields here are $ x_0 $, $ \zeta $, $ n^l $ and $ \xi^l $.
	In the absence of deformation, the fermionic auxiliary variables $ \lambda_{L,R} $  are responsible for 
	the CP($N-1$) constraints \eqref{constr}.
	In the \ntwoo theory, however, the superpotential $ \mathcal{W}_{1+1} $ prevents these constraints to be satisfied
	 for the {\it right-handed} coordinates $ \xir $, $ \bxir $:
\[
	\nbar{}_l\, \xi_L^l ~~=~~ 0\,, \qquad\qquad  
	\ov{\xi}{}_{lR}\, n^l ~~=~~ \delta\, \frac{m_W}{\sqrt{2}}\, \zeta_R~.
\]
	For practical reasons, however, it is convenient to restore CP($N-1$) constraints by a shift
\begin{align}
\notag
	\bxir' ~~=~~ \bxir ~-~ \frac{m_W}{\sqrt{2}}\, \delta \cdot \zeta_R\ov{n} \,,\\[2mm]
\label{shift}
	\xi_R' ~~=~~ \xi_R ~-~ \frac{m_W}{\sqrt{2}}\, \ov{\delta} \cdot \bzr n~.
\end{align}
	With this redefinition, and all auxiliary fields excluded, the sigma model takes the form
\begin{align}
\notag
S_{1+1}^{(0,2)} ~~=&~~
	\int d^2 x 
\Biggl\lgroup
	2\pi\xi \lgr \frac 1 2 \left (\p x_0\right)^2 ~+~
		      \frac 1 2 \bzl\, i\p_R\, \zl ~+~
			\frac 1 2 \bzr\, i\p_L\, \zr
		\rgr
	 \\[2mm]
\label{02_unnorm}
&
	+2\beta\, \Bigg\{
	\left|\p n\right|^2 ~+~ \left(\ov{n}\p_k n\right)^2 ~+~
	\bxir \, i\p_L \, \xir  ~+~ \bxil \, i\p_R \, \xil \\[2mm]
\notag
&
	~~~~
	-~
	i \left(\ov{n}\p_L n\right)\, \bxir \xir ~-~ 
	i \left(\ov{n}\p_R n\right)\, \bxil \xil 
	\\[2mm]
\notag
&
	~~~~
	~+~
	\frac{m_W/\sqrt{2}} { \sqrt{ 1~+~2|\delta|^2 } }
	\lgr \delta\, (i\p_L \ov{n}) \xir\zr ~+~ 
             \ov{\delta}\, \bxir (i\p_L n)\bzr 
	\rgr \\[2mm]
\notag
&
	~~~~
	~+~ \frac{1}{1 + 2|\delta|^2}\, \bxil\xir \bxir\xil 
	~-~ \bxil\xil\bxir\xir
	~+~
	\frac{m_W^2}{2}\,\frac{|\delta|^2}{1 ~+~ 2|\delta|^2}\,
		\bxil\xil\bzr\zr \Bigg\} 
\Biggr\rgroup~.
\end{align}
	While removing the unpleasant right-hand side in the CP($N-1$) constraints, the
	substitution \eqref{shift} introduces interaction between the superorientational
	and supertranslational moduli, {\it e.g.} the so-called ``bifermionic mixing term''
	in the fourth line of Eq.~\eqref{02_unnorm}.
	The latter is seen most explicitly if one rescales the translational variables,
\[
	\zr ~~=~~ \frac {\zr'}  {{m_W}/{2}}~, \qquad {\rm etc.}~
\]
	with the same substitution for $ x_0 $ and $ \zl $.
	Then one obtains 
\begin{align}
\notag
S_{1+1}^{(0,2)} ~~=~~ 2\beta
	\int & d^2 x 
\lgr
	\bzr\, i\p_L\, \zr ~~+~~ \dots 
\right.
	\\[2mm]
\label{world02}
	&
	\;\;
	+~~
	\left|\p n\right|^2 ~~+~~ \left(\ov{n}\p_k n\right)^2 ~~+~~
	\bxir \, i\p_L \, \xir  ~~+~~ \bxil \, i\p_R \, \xil 
	\\[2mm]
\notag
	&
	\;\;
	-~~
	i \left(\ov{n}\p_L n\right)\, \bxir \xir ~~-~~ 
	i \left(\ov{n}\p_R n\right)\, \bxil \xil  
	\\[2mm]
\notag
	&
	\;\;
	+~~
	\tgamma\, (i\p_L\nbar) \xir\zr ~~+~~ \ov{\tgamma}\, \bxir (i\p_L n) \bzr ~~+~~
	|\tgamma|^2\, \bxil\xil \bzr\zr  
	\\[2mm]
\notag
	&
	\;\;
\left.
	+~~ 
	\left( 1 \;-\; |\tgamma|^2 \right)\, \bxil\xir \bxir\xil  
	~~-~~ \bxil\xil \bxir\xir
\rgr ,
\end{align}
	where the ellipses stand  for the decoupled part of the theory,
\[
	\dots ~~=~~ \left(\p x_0\right)^2 ~~+~~ \bzl\, i\p_R\, \zl ~,
\]
	and the coefficient $ \tgamma $ in front of the bifermionic term is related to the superpotential parameter $ \delta $ via
\[
	\tgamma ~~=~~ \frac { \sqrt{2}\,\delta } { \sqrt{ 1 +  2 |\delta|^2 } }~.
\]
	In this paper the heterotic deformation parameter $ \tgamma $ is connected with the analogous parameter
	$ \gamma $ introduced in \cite{SYhet} by the following relation,
\[
	\tgamma ~~=~~ \sqrt{2/\beta}\, \gamma\,.
\]
	The theory \eqref{world02} is invariant under the \ntwoo supersymmetry transformations
\begin{align*}
	&
	\delta n ~~=~~ \sqrt{2}\, \epsilon_R \xil \,, \\[2mm]
	&
	 \delta\ov{n} ~~=~~ \sqrt{2}\, \ov{\epsilon_R \xi}{}_L\,,
	\\[2mm]
	&
	\delta\xil ~~=~~ i\sqrt{2}\, \ov{\epsilon}{}_R \p_L n ~~-~~ 
			\sqrt{2}\, \ov{\epsilon}{}_R \bxil\xil \cdot n ~~-~~
			i\sqrt{2}\, \ov{\epsilon}{}_R \left( \ov{n}\p_L n \right) \cdot n \,,\\[2mm]
	&
	\delta\bxil ~~=~~ i\sqrt{2}\epsilon_R \p_L \ov{n}  ~~+~~
			\sqrt{2}\,\epsilon_R \bxil\xil \cdot \ov{n} ~~+~~
			i\sqrt{2}\, \epsilon_R \left( \ov{n} \p_L n \right) \cdot \ov{n}\,, \\[2mm]
	& 
	\delta\xir ~~=~~ - \sqrt{2}\, \ov{\epsilon}{}_R \bxil\xir \cdot n 
		~~-~~ \sqrt{2}\, \ov{\tgamma}\epsilon_R\, \xil \bzr \,,\\[2mm]
	&
	\delta\bxir  ~~=~~ \sqrt{2}\, \epsilon_R \bxir \xil \cdot \ov{n} 
		~~-~~ \sqrt{2}\, \tgamma\ov{\epsilon}{}_R\, \bxil\zr \,,\\[2mm]
	&
	\delta\zeta_R ~~=~~ -\, \sqrt{2}\, \ov{\tgamma}\epsilon_R \cdot \bxir\xil \,,\\[2mm]
	&
	\delta\ov{\zeta}{}_R ~~=~~ \sqrt{2}\, \tgamma \ov{\epsilon}{}_R \bxil\xir~,
\end{align*}
	which represent the right-handed half of the \ntwot supersymmetry \eqref{susy_ntwot} 
	deformed by the parameter $ \tgamma $.
	It is straightforward to see that these supertransformations preserve the CP($N-1$) constraints
\[
	\ov{n}{}_l n^l ~~=~~ 1~, \qquad\qquad  \ov{n}{}_l\xi_\alpha^l ~~=~~ \ov{\xi}{}_{\alpha l} n^l ~~=~~ 0~,
		\qquad\qquad\qquad  \alpha ~=~ R,L~.
\]
The normalized form of the \CPC action \eqref{world02} can be readily transformed into yet another
	formulation of the model, discovered in \cite{SYhet} --- the geometric formulation --- which we  will briefly describe here.

\newcommand{\bi}{{\bar \imath}}
\newcommand{\bj}{{\bar \jmath}}
\newcommand{\bk}{{\bar k}}
\newcommand{\bl}{{\bar l}}
\newcommand{\bm}{{\bar m}}
	Geometric formulation of \CPC model is based on the K\"{a}hler formulation of the CP($N-1$) supersymmetric
	sigma model.
	One has two sets of $ N - 1 $ (anti)chiral superfields $ \Phi^i $ and $ \ov{\Phi}{}^\bj $, 
	$ i, \bj = 1,..., N-1 $, the lowest components $ \phi^i $, $ \ov{\phi}^\bj $ of which parametrize the target K\"{a}hler
	manifold.
The Lagrangian of the CP($N-1$) model is given by the following sigma model
\[
	\mc{L} ~~=~~ \int\, d^4\theta\, K(\Phi,\ov{\Phi}) ~~=~~ g_{i\bj}\,\p_\mu \phi^i \p_\mu\ov{\phi}{}^\bj
		~+~ \frac{1}{2}\, g_{i\bj}\, \psi^i\, i\overleftrightarrow{\slashed{\nabla}} \ov{\psi}{}^\bj 
		~+~ \frac{1}{4}\, R_{ij\bk\bl}\, \psi^i\psi^j \ov{\psi}{}^\bk \ov{\psi}{}^\bl~,
\]
	where $ K(\phi,\ov{\phi}) $ is the K\"ahler potential, 
	$ g_{i\bj} $ is its K\"ahler metric
\[
	g_{i\bj} ~~=~~ \frac{\p^2 K}{\p\phi^i\,{\p\ov{\phi}{}^\bj}}~,
	\qquad\qquad
	g^{i\bk} ~~=~~ \left(g^{-1}\right)^{\bk i}~,
\]
	$ \nabla_\mu $ the covariant derivative,
\begin{align*}
	(\nabla_\mu \ov{\psi})^\bj & ~~=~~ \left\{ \p_\mu \delta^\bj_{\ \bm} ~+~
						\Gamma_{\bm\bk}^\bj\, \p_\mu(\ov{\phi}{}^\bk) \right\} \ov{\psi}{}^\bm~,
	& \Gamma^\bi_{\bk\bl} & ~~=~~ g^{m\bi}\,\p_\bl\, g_{m\bk}~,
	\\[3mm]
	(\psi \overleftarrow{\nabla}{}_\mu)^i & ~~=~~ 
			\psi^m \left\{ \overleftarrow{\p}{}_\mu\delta^i_{\ m} ~+~
						\Gamma^i_{mk}\, \p_\mu(\phi^k) \right\}~,
	& \Gamma^i_{kl} & ~~=~~ g^{i\bm}\, \p_l\, g_{k\bm}~,
\end{align*}
	and $ R_{ij\bk\bl} $ the Riemann tensor 
\[
	R_{ij\bk\bl} ~~=~~ \p_i\,\p_\bk\, g_{j\bl} ~-~ g^{m\bm}\; \p_i\, g_{j\bm}\, \p_\bk g_{m\bl}~.
\]
	For the CP($N-1$) model one chooses the K\"ahler potential
\[
	K(\Phi, \ov{\Phi}) ~~=~~ \ln \lgr 1 ~+~ \ov{\Phi}{}^\bj \delta_{\bj i} \Phi^i \rgr
\]
	which corresponds to the Fubini--Study metric,
\[
	g_{i\bj} ~~=~~ \frac{1}{\chi}\lgr  \delta_{i\bj} ~-~ \frac{1}{\chi}
				  \delta_{i\bi}\,\ov{\phi}^\bi\, \delta_{j\bj}\,\phi^j \rgr,
	\qquad\qquad \text{where~~}
	\chi ~~=~~ 1 ~+~ \ov{\phi}{}^\bj \delta_{\bj i} \phi^i~.
\]
	In this case,
\[
	\Gamma^\bi_{\bk\bl} ~~=~~ -\, \frac{\delta^\bi_{\ (\bk} \delta_{\bl) i}\, \phi^i}{\chi}\,,  
	\qquad\qquad 
	\Gamma^i_{kl} ~~=~~ -\, \frac{\delta^i_{\ (k} \delta_{l)\bi}\,\ov{\phi}{}^\bi}{\chi}\,,
\]
	and the Riemann tensor takes the form
\[
	R_{ij\bk\bl} ~~=~~ -\,g_{i(\bk}\,g_{\bl)j}~.
\]
	
	As was shown in \cite{SYhet},  the \ntwoo deformation of the CP($N-1$) model can be achieved by
	introduction of the right-handed supertranslational modulus $ \zeta_R $ via a ``right-handed'' 
	supermultiplet $ \mc{B} $,
\begin{align*}
	\mc{B} & ~~=~~ \lgr \zr ~+~ \sqrt{2}\,\theta_R\mc{F} \rgr \ov{\theta}{}_L~, \\[2mm]
	\ov{\mc{B}} & ~~=~~ \theta_L \lgr \bzr ~+~ \sqrt{2}\, \ov{\theta}{}_R \ov{\mc{F}} \rgr.
\end{align*}
	The latter expressions describe superfields covariant only under the right-handed supersymmetry, 
	while explicitly breaking the left-handed one.
	In a sense, $ \mc{B} $ is the analogue of the \ntwoo supermultiplet $\Xi$ in the two-dimensional 
	superfield formalism \cite{Edalati} ---
	the supermultiplet containing only one physical field, which is the supertranslational
	fermionic variable.
	One then constructs the action 
\beq
\label{exte}
	\mc{L}_{(0,2)} ~~=~~ \frac{2}{g_0^2}\, \int\, d^4\theta\, \lgr K(\Phi,\ov{\Phi}) 
		~-~ 2\, \ov{\mc{B}}\,\mc{B}  ~+~  \sqrt{2}\,\tgamma\,\mc{B}\,K  
					     ~+~ \sqrt{2}\,\ov{\tgamma}\,\ov{\mc{B}}\,\ov{K} \rgr,
\eeq
	which respects the invariance on the target space CP($N-1$).
	Here
\[
	\frac{2}{g_0^2} ~~=~~ 2\beta
\]
	defines the coupling constant of the sigma model.
	The second term in \eqref{exte} generates the kinetic term for $ \zr $, while the last two terms 
	are responsible for the mixing between $ \zr $ and $ \xi_{R,L} $.
	Explicitly, one has,
\begin{align}
\notag
	\frac{\mc{L}_{(0,2)}}{2\beta}
                    & ~~=~~  \bzr\, i\p_L\, \zr 
			~+~ g_{i\bj}\,\p_\mu \phi^i \p_\mu\ov{\phi}{}^\bj
			~+~ \frac{1}{2}\, g_{i\bj}\, \psi^i\, i\overleftrightarrow{\slashed{\nabla}} \ov{\psi}{}^\bj 
	\\[2mm]
\label{cpn-1g}
			& 
			~~+~~ \tgamma\, g_{i\bj}\, (i \p_L \ov{\phi}{}^\bj)\, \psi_R^i\, \zr
			~+~ \ov{\tgamma}\, g_{i\bj}\, \ov{\psi}{}_R^\bj (i \p_L \phi^i)\, \bzr
			~+~ |\tgamma|^2\, \bzr\,\zr \cdot ( g_{i\bj}\, \ov{\psi}{}_L^\bj\, \psi_L^i )
	\\[2mm]
\notag
			& 
			~~+~~ (1 \!-\! |\tgamma|^2)\, (g_{i\bk}\, \ov{\psi}{}_R^\bk\, \psi_L^i)\,
						     (g_{j\bl}\, \ov{\psi}{}_L^\bl\, \psi_R^j)
			~-~ (g_{i\bk}\, \ov{\psi}{}_R^\bk\, \psi_R^i)\, (g_{j\bl}\, \ov{\psi}{}_L^\bl\, \psi_L^j)~.
\end{align}
	To be able to match the Lagrangian in this formula to that of Ref.~\cite{SYhet},
	one needs express $ \tgamma $ in terms of $ \gamma $ via
$ \tgamma = \sqrt{2}g_0\gamma $ and normalize the kinetic term for $ \zr $ 
	canonically.
	
	The geometric form \eqref{cpn-1g} can be related to the ``gauge'' form \eqref{world02}
	via the following stereographic projection
\begin{align*}
	n^i & ~~=~~ \frac{\phi^i}{\sqrt{\chi}}~,
	& 
	\ov{n}_\bi & ~~=~~ \frac{\ov{\phi}{}^\bi}{\sqrt{\chi}}~,
\\[2mm]
	n^N & ~~=~~ \frac{1}{\sqrt{\chi}}~,
	& n^N & ~~\in~~ \mc{R}~,
\\[2mm]
	\xi^i & ~~=~~ \frac{1}{\sqrt{\chi}} \lgr \psi^i ~-~ \frac{(\ov{\phi}\psi)}{\chi}\,\phi^i \rgr,
	& 
	\ov{\xi}{}_\bi & ~~=~~ \frac{1}{\sqrt{\chi}} 
					\lgr \ov{\psi}{}^\bi ~-~ \frac{(\ov{\psi}\phi)}{\chi}\, \ov{\phi}{}^\bi \rgr,
\\[2mm]
	\xi^N & ~~=~~ -\, \frac{(\ov{\phi} \psi)}{\chi^{3/2}}~,
	&
	\ov{\xi}{}_N & ~~=~~ -\, \frac{(\ov{\psi} \phi)}{\chi^{3/2}}~,
\end{align*}
	where $	i,\, \bi ~=~ 1, ..., N-1 $ and we shortcut the contractions 
	$ (\ov{\psi} \phi) ~=~ \delta_{i\bj}\, \ov{\psi}{}^\bj \phi^i $.
	Here we chose $ n^N $ to be real given an overall phase freedom of the CP($N-1$) variables $ n^l $.

%
%
\section{Worldsheet \boldmath\ntwoo Theory from the Microscopic \boldmath\none Theory}
\label{WS}
\setcounter{equation}{0}

	As we mentioned before, the bosonic string solution in the \none case remains the same,
	which helps in finding the effective world sheet theory on the string. 
	At the technical level, we got convinced in the previous section that the difference
	between the \ntwoo and \ntwot theories is most evidently exemplified by the bifermionic cross-term,
	see the fourth line in Eq.~\eqref{world02}.
	Apart from that term and some quartic terms proportional to $|\tgamma|^2$, the answer for effective 
	world sheet theory is given by Eq.~\eqref{str_ntwot}.
	As for the bifermionic term, it arose from the mixing of the kinetic terms of 
	superorientational and supertranslational moduli via substitution \eqref{shift}, and it is therefore 
	natural to look for the presence of the former term in the effective theory by using fermionic zero modes 
	in the kinetic part of the microscopic theory. 
	Had we not done the substitution \eqref{shift}, our goal would have been finding the quartic fermionic terms,
	a much more difficult task. 

	One difficulty here is that the theory on the string possesses the light modes with masses \eqref{mUm}, \eqref{mNm},
	which become massless in the $ \mu \to \infty $ limit.
	This limit one wants to take to proceed to \none SQCD, which possesses a Higgs branch.
	However, as long as one keeps $ \mu $ finite, the result for the bifermionic mixing can be calculated.
	The presence of the light modes will be then seen as the long $ 1/r $ tails of the zero modes,
	divergent when $ \mu \to \infty $.

	Another difficulty now is that half of supersymmetry which was used for calculating the fermionic zero modes, is lost.
	That brings one to the necessity of solving the Dirac equations, which was done in \cite{SYhet} for the SU(2) theory.
	
	The other half of supersymmetry is still there, and can be used to obtain the left-handed zero modes. 
	Since the bosonic string solution did not undergo any change when \ntwo was broken, the corresponding zero modes
	must be the same as in the \ntwo theory.
	For the supertranslational modes they are those proportional to $ \zeta_L $, and for the superorientational
	modes --- those proportional to $ \xi_L $.
	The zero modes proportional to $ \zeta_R $ and $ \xi_R $ have changed and must be obtained from the Dirac equations. 

	Let us start with the supertranslational modes.
	It is easy to guess a good {\it ansatz} for the latter,
\begin{align}
\notag
	\lambda^{22\ \rm U(1)} & ~~=~~ \loU\, \zeta_R ~+~ \llU\, \frac{x^1 + i x^2}{r} \ov{\zeta}{}_R \,,
	\\[2mm]
\notag
	\lambda^{22\ \text{SU($N$)}} & ~~=~~ \lgr  \loN\, \zeta_R ~+~ \llN\, \frac{x^1 + i x^2}{r} \ov{\zeta}{}_R \rgr
					( n\nbar ~-~ 1/N )\,,
	\\[2mm]
\label{ftprofile}
	\ov{\wt{\psi}}{}_{\dot{1}} & ~~=~~ \frac{1}{2} \frac{x^1 - i x^2}{r}
				\lgr  \poU ~+~ N (n\nbar ~-~ 1/N) \poN \rgr \zeta_R 
				\,,\\[2mm]
\notag
				   & 
				~~+~~ \frac{1}{2} \lgr  \plU  ~+~ N (n\nbar ~-~ 1/N) \plN \rgr  \ov{\zeta}{}_R
	~.
\end{align}
	Here $ \lambda^{\rm U(1)}_{1,2} $, $ \lambda^\text{SU($N$)}_{1,2} $, $ \psi^{\rm U(1)}_{1,2} $ and
	$ \psi^\text{SU($N$)}_{1,2} $ are the profile functions to be determined.
	This {\it ansatz} is very natural from the standpoint of our smooth \none deformation. 
	When the parameters $ \mu_{1,2} $ are very small, the profiles with the subscript ``0'' become the \ntwo zero modes,
	since they are proportional to the unbarred $ \zeta_R $, and therefore can be taken from Eq.~\eqref{N2_strans}.
	The profiles with the subscript ``1'' become deformations, as they are proportional to $ \bzr $.
	This interpretation, however, is only valid for very small $ \mu $, when, as  will be seen, the deformations
	are indeed directly proportional to $ \mu $.
	At large $ \mu $, the ``1''-profiles become large and lose their meaning as deformations, accompanied by the fact that 
	the ``0''-profiles will no longer be given by the \ntwo equations \eqref{N2_strans}.

	We substitute now the {\it ansatz} \eqref{ftprofile} into the Dirac equations \eqref{dirac} to obtain the following
	equations for the profiles:
\begin{align}
\notag
&
	-\, \p_r \loU ~+~ \frac{i g_1^2}{4\sqrt{2}} 
			\lgr \poU (\phi_1 + \phi_2) ~+~ (N-1) \poN (\phi_1 - \phi_2) \rgr  \\[2mm]
\notag
	&\qquad\qquad\qquad\qquad\qquad\qquad\qquad\qquad\qquad\qquad\qquad
				~+~ g_1^2 \sqrt{\frac{N}{2}} \mu_1 \llU    ~~=~~ 0\,,
	\\[2mm]
\notag
&
	-\, \p_r \llU ~-~ \frac{1}{r}\llU 
	~+~ \frac{i g_1^2}{4\sqrt{2}} 
	    \lgr \plU (\phi_1 + \phi_2) ~+~ (N-1) \plN (\phi_1 - \phi_2) \rgr \\[2mm]
\notag
	&\qquad\qquad\qquad\qquad\qquad\qquad\qquad\qquad\qquad\qquad\qquad
	~+~ g_1^2 \sqrt{\frac{N}{2}} \mu_1 \loU ~~=~~ 0\,,
	\\[2mm]
\label{lambdaeqs}
&
	-\, \p_r \loN ~+~ 
	\frac{i g_2^2}{2\sqrt{2}}
		\lgr \poU (\phi_1 - \phi_2) ~+~ \poN ( (N-1) \phi_1 + \phi_2 ) \rgr \\[2mm]
\notag
	&\qquad\qquad\qquad\qquad\qquad\qquad\qquad\qquad\qquad\qquad\qquad
	~+~ g_2^2\, \mu_2 \llN ~~=~~ 0\,,
	\\[2mm]
\notag
&
	-\, \p_r \llN ~-~ \frac{1}{r}\llN
	~+~ \frac{i g_2^2}{2\sqrt{2}} 
		\lgr \plU (\phi_1 - \phi_2) ~+~ \plN ( (N-1) \phi_1 + \phi_2 ) \rgr \\[2mm]
\notag
	&\qquad\qquad\qquad\qquad\qquad\qquad\qquad\qquad\qquad\qquad\qquad
	~+~ g_2^2\, \mu_2 \loN ~~=~~ 0\,,
\end{align}
	for the gauginos, and 
\begin{align}
\notag
&
	\p_r \poU ~+~ \frac{1}{r} \poU ~-~ \frac{1}{Nr}f \poU ~-~ \frac{N-1}{Nr} f_N \poN 
	 \\[2mm]
&\notag
\qquad\qquad\qquad
	+\,i\, \frac{2\sqrt{2}}{N} 
		\lgr  \loU (\phi_1 + (N-1)\phi_2) ~+~ \frac{N-1}{N} \loN (\phi_1 - \phi_2) \rgr 
		~~=~~ 0\,,
	\\[2mm]
&
\notag
	\p_r \plU ~-~ \frac{1}{Nr}f \plU ~-~ \frac{N-1}{Nr} f_N \plN 
	 \\[2mm]
&\notag
\qquad\qquad\qquad
	+\,i\, \frac{2\sqrt{2}}{N}
		\lgr \llU (\phi_1 + (N-1) \phi_2) ~+~ \frac{N-1}{N} \llN (\phi_1 - \phi_2) \rgr
		~~=~~ 0\,,
	\\[2mm]
&
\label{psieqs}
	\p_r \poN ~+~ \frac{1}{r} \poN ~-~ \frac{1}{Nr} (f + (N-2)f_N) \poN ~-~
			\frac{1}{Nr} f_N \poU 
	 \\[2mm]
&\notag
\qquad\qquad\qquad
	+\, i\, \frac{2\sqrt{2}}{N} 
		\lgr \loU (\phi_1 - \phi_2) ~+~ \frac{1}{N} \loN ((N-1)\phi_1 + \phi_2) \rgr
		~~=~~ 0\,,
	\\[2mm]
&
\notag
	\p_r \plN ~-~ \frac{1}{Nr} (f + (N-2)f_N) \plN - \frac{1}{Nr} f_N \plU 
	   \\[2mm]
&\notag
\qquad\qquad\qquad
	+\,i\, \frac{2\sqrt{2}}{N}
		\lgr \llU (\phi_1 - \phi_2) ~+~ \frac{1}{N} \llN ((N-1)\phi_1 + \phi_2) \rgr
		~~=~~ 0~
\end{align}
	for the quarks.
	These equations are easy to solve in either small-$ \mu $ or large-$ \mu $ limit.

	Now we will guess a similar {\it ansatz} for the right-handed superorientational modes, where it
	is quite natural to write,
\begin{align}
\notag
	\lambda^{22\ \text{SU($N$)}} & ~~=~~ 2\, \frac{x^1 + ix^2}{r}\, \lambda_+(r) \; \xi_R\ov{n}
				~~+~~  2\, \lambda_-(r)\; n\ov{\xi}{}_R\,,
	\\[2mm]
\label{fprofile}
	\ov{\wt{\psi}}{}_{\dot{1}} & ~~=~~ 2\, \psi_+(r)\; \xi_R \ov{n} 
				~~+~~  2\, \frac{x^1 - i x^2}{r}\, \psi_-(r)\; n\ov{\xi}{}_R~.
\end{align}
	Here $ \lambda_+(r) $ and $ \psi_+(r) $ represent the ``undeformed'' profile functions in the sense
	explained above, while $ \lambda_-(r) $ and $ \psi_-(r) $ are the ``perturbations'' due to supersymmetry
	breaking.

	Substituting \eqref{fprofile} into the Dirac equations \eqref{dirac}, one obtains
\begin{align}
\notag
&
	\p_r \psi_+ ~-~ \frac{1}{Nr} (f-f_N)\, \psi_+ ~+~ i\,\sqrt{2}\phi_1\,\lambda_+ ~~=~~ 0\,,
	\\[2mm]
\notag
	-\, & \p_r\lambda_+ ~-~ \frac{1}{r}\lambda_+ ~+~ \frac{f_N}{r}\lambda_+ 
		~+~ i\,\frac{g_2^2}{\sqrt{2}}\phi_1\, \psi_+ ~+~ \mu_2 g_2^2\, \lambda_-  ~~=~~ 0\,,
	\\[2mm]
\label{fermeqs}
&
	\p_r \psi_- ~+~ \frac{1}{r}\, \psi_- ~-~ \frac{1}{Nr}(f + (N-1)f_N)\, \psi_- 
							~+~ i\,\sqrt{2}\phi_2\, \lambda_- ~~=~~ 0\,,
	\\[2mm]
\notag
	-\, & \p_r\lambda_- ~-~ \frac{f_N}{r}\lambda_- ~+~ i\,\frac{g_2^2}{\sqrt{2}}\phi_2\, \psi_- 
								~+~ \mu_2 g_2^2\, \lambda_+ ~~=~~ 0
	~.
\end{align}
	The supersymmetry breaking parameter $ \mu_1 $ does not enter these equations since the  U(1) 
	multiplet does not develop orientational, {\it i.e.} non-Abelian zero modes by definition.

\subsection{Small-\boldmath{$\mu$} limit}
As explained above, when $ \mu $ is very small, the Dirac equations can be solved perturbatively. 
We choose a particular relation between the parameters of the Abelian and non-Abelian deformations $ \mu_1 $ 
and $ \mu_2 $,
\beq
\label{mueq}
	g_1^2 \sqrt{\frac{N}{2}}\, \mu_1 ~~=~~ g_2^2 \mu_2  \qquad\qquad \Longleftrightarrow \qquad\qquad 
		\mUp  ~~=~~ \mNp~,
\eeq
which turns out to be convenient.
The perturbation series run in powers of $ \mu^2 $.
The ``0''-profiles run in even powers of $ \mu $ and at the leading order constitute the ``\ntwon''-supersymmetric zero modes, 
while the ``1''-profiles run in odd powers of $ \mu $ and at the leading order are the ``deformations''.
For the supertranslational modes, the {\it ansatz} \eqref{ftprofile} determines the ``0''-profiles
from the  \ntwo zero modes, Eq.~\eqref{N2_strans}:
\begin{align}
\notag
	\loU & ~~=~~ i\, \frac{g_1^2}{2}\, \lgr (N-1) \phi_2^2 ~+~ \phi_1^2 ~-~ N\xi \rgr  ~+~ O(\mu^2)\,, 
	\\[2mm]
\notag
	\loN & ~~=~~ i\, g_2^2 \lgr \phi_1^2 ~-~ \phi_2^2 \rgr ~+~ O(\mu^2)\,,
	\\[2mm]
\label{tzeroorder}
	\poU & ~~=~~ \frac{4\sqrt{2}}{N^2 r} \lgr \phi_1\, (f + (N-1) f_N) ~+~ (N-1)\, \phi_2\, (f-f_N) \rgr ~+~ O(\mu^2)\,,
	\\[2mm]
\notag
	\poN & ~~=~~ \frac{4\sqrt{2}}{N^2 r} \lgr \phi_1\, (f + (N-1) f_N) ~-~ \phi_2\, (f-f_N) \rgr ~+~ O(\mu^2)
	~.
\end{align}
	Inspecting Eqs.~\eqref{lambdaeqs} and \eqref{psieqs} and dropping all terms proportional to $ O(\mu^2) $ one 
	immediately observes that the first-order profiles obey  differential equations similar
	to those of the zero-order 
	profiles, and thus are proportional to the latter,
\begin{align}
\notag
	& \plU ~~=~~ \frac{g_2^2\mu_2}{2}\, r \, \poU ~+~ O(\mu^3)\,,		& \plN &~~=~~ \frac{g_2^2\mu_2}{2}\, r\, \poN ~+~ O(\mu^3)  \,,
	\\[2mm]
\label{tfirstorder}
	& \llU ~~=~~ \frac{g_2^2\mu_2}{2}\, r \, \loU ~+~ O(\mu^3)\,,		& \llN &~~=~~ \frac{g_2^2\mu_2}{2}\, r\, \loN ~+~ O(\mu^3)~.
\end{align}
As for the superorientational zero modes, the zero-order profiles $ \lambda_+ $ and $ \psi_+ $ are taken from
	\eqref{N2_sorient},
\begin{align}
\notag
 	\lambda_+(r) & ~~=~~ -\, \frac{i}{\sqrt{2}} \frac{f_N}{r} \frac{\phi_1}{\phi_2}  ~+~ 
	O(\mu_2^2) \,,\\[2mm]
\label{zeroorder}
	\psi_+(r) & ~~=~~ -\, \frac{\phi_1^2 ~-~ \phi_2^2}{2\phi_2} ~+~ O(\mu_2^2)~.
\end{align}
	The leading-order contributions to the $ \psi_- $ and $ \lambda_- $ profile functions can be shown to be 
\begin{align}
\notag
	\psi_- & ~~=~~ -\, \mu_2 g_2^2 \frac{r}{4\phi_1} \left( \phi_1^2 ~-~ \phi_2^2 \right)  ~+~ O(\mu^3)\,,
	\\[2mm]
\label{firstorder}
	\lambda_- & ~~=~~ -\, \mu_2 g_2^2 \frac{i}{2\sqrt{2}} \lgr (f_N - 1) \frac{\phi_2}{\phi_1} ~+~ \frac{\phi_1}{\phi_2} \rgr ~+~ O(\mu^3)~.
\end{align}
	One easily checks that the above solutions behave well at the origin and at infinity.

\subsection{Large-\boldmath{$\mu$} limit}
	When $ \mu $ is large, one cannot treat the zero modes problem perturbatively.
	Simplifying arguments are needed to be brought up.
	One obvious remark is that at large $ \mu $ the adjoint fields become heavy,
	and effectively stop propagating.
	In terms of the profile functions it means that the kinetic terms for $ \lambda^2 $'s 
	can be dropped out from equations \eqref{lambdaeqs} and \eqref{fermeqs}.
	Then $ \lambda^2 $ can be resolved and completely excluded from the equations.
	
	Another argument is that we are only interested in the long-distance behavior
	of the zero modes, which presumably will be divergent in $ \mu $, in the sense that the
	zero modes will cease to be normalizable at $ \mu \to \infty $.
	At large distances the string profile functions are very close to their asymptotic values \eqref{boundary},
	which will allow us to significantly simplify the equations.
	In particular, the gauge profile functions vanish at infinity.
	On the other hand, {\it e.g.} in Eqs.~\eqref{psieqs}, it is the functions $ f $ and $ f_N $ 
	which bind the Abelian and non-Abelian profile functions together.
	With $ f $ and $ f_N $ neglected ({\it cf.} Eq.~\eqref{boundary}), this binding is lost and 
	the solutions will turn out to be independently normalized.
	To restore their mutual normalization we will need to step back to lower distances and consider
	$ f $'s nonvanishing.
	We therefore deal with two cases, large $ r $ and intermediate $ r $, in turn, for supertranslational
	and superorientational modes.

\subsubsection{Supertranslational zero modes}
	{\flushleft{\it Large-$r$ domain: $ r \gg 1/(g\sqrt{\xi}) $}.}\\[2mm]
	Dropping the kinetic terms for $ \lambda $'s in \eqref{lambdaeqs} and setting the string
	profiles $ \phi_1 $, $ \phi_2 $, $ f $ and $ f_N $ to their asymptotic values, one has
	from Eqs.~\eqref{lambdaeqs}, \eqref{psieqs}
\begin{align}
\notag
& 
	\frac{i}{2\sqrt{2}} \sqrt{\xi} \poU ~+~ \mu_1 \sqrt{\frac{N}{2}}\, \llU ~=~ 0\,,   
&&
	\p_r \,\poU ~+~ \frac{1}{r}\,\poU ~+~ i\, 2\sqrt{2}\sqrt{\xi}\, \loU ~=~ 0 \,,
\\[2mm]
\notag
& 
	\frac{i}{2\sqrt{2}} \sqrt{\xi} \plU ~+~ \mu_1 \sqrt{\frac{N}{2}}\, \loU ~=~ 0\,,      
&&
	\p_r \, \plU ~+~ i\,2\sqrt{2}\sqrt{\xi}\,\llU ~~=~~ 0\,,
\\[2mm]
\label{strans_red}
&
	\frac{i}{2\sqrt{2}} N\sqrt{\xi} \poN ~+~ \mu_2 \llN ~=~ 0\,,     
&&
	\p_r \,\poN ~+~ \frac{1}{r}\, \poN  \\[2mm]
\notag &&& \qquad\qquad
	~+~ i\,\frac{2\sqrt{2}}{N} \sqrt{\xi}\, \loN ~=~ 0\,,
\\[2mm]
\notag
&
	\frac{i}{2\sqrt{2}} N\sqrt{\xi} \plN ~+~ \mu_2 \loN ~=~ 0\,,   
&&
	\p_r \,\plN ~+~ i\,\frac{2\sqrt{2}}{N}\sqrt{\xi} \llN ~=~ 0~.
\end{align}
	Keeping the notation for the light masses 
\[
	\mUm ~~=~~ \sqrt{\frac{N}{2}}\, \frac{\xi}{\mu_1}~,  \qquad\qquad \mNm ~~=~~ \frac{\xi}{\mu_2}~,
\]
	we resolve the heavy gauginos
\begin{align}
\notag
	\loU & ~=~ -\, \frac{i}{2\sqrt{2}}\, \frac{2}{N}\, \frac{\mUm}{\sqrt{\xi}}\, \plU \,,
&
	\llU & ~=~ -\, \frac{i}{2\sqrt{2}}\, \frac{2}{N}\, \frac{\mUm}{\sqrt{\xi}}\, \poU \,,\\[2mm]
\label{heavy_gauginos}
	\loN & ~=~ -\, \frac{i}{2\sqrt{2}}\, N\, \frac{\mNm}{\sqrt{\xi}}\, \plN\,,
&
	\llN & ~=~ -\, \frac{i}{2\sqrt{2}}\, N\, \frac{\mNm}{\sqrt{\xi}}\, \poN ~,
\end{align}
	and substitute them back into \eqref{strans_red},
\begin{align*}
	& \p_r\, \plU ~+~ \frac{2}{N}\,\mUm\,\poU ~~=~~ 0\,, \\[2mm]
	& \p_r\, \poU ~+~ \frac{1}{r}\,\poU ~+~ \frac{2}{N}\,\mUm\,\plU ~~=~~ 0\,, \\[2mm]
	& \p_r\, \plN ~+~ \mNm\,\poN ~~=~~ 0 \,,\\[2mm]
	& \p_r\, \poN ~+~ \frac{1}{r}\,\poN ~+~ \mNm\,\plN ~~=~~ 0~.
\end{align*}
	From these expressions, one obtains the second-order equations for $ \plU $ and $ \plN $,
\begin{align}
\notag
	& \p_r^2\, \plU  ~~+~~ \frac{1}{r}\p_r\, \plU ~-~ \frac{4}{N^2}\,(\mUm)^2\, \plU  ~~=~~ 0~, 
	\\[2mm]
\label{free_strans}
	& \p_r^2\, \plN  ~~+~~ \frac{1}{r}\p_r\, \plN ~-~ (\mNm)^2\, \plN ~~=~~ 0~.
\end{align}
	The solutions are given in terms of  the McDonald function $ K_0(r) $, namely,
\begin{align*}
	\plU & ~=~ \frac{2}{N}\,C\,\mUm\,\sqrt{\xi}\;K_0\left(\frac{2}{N}\mUm\,r\right)  ,
&
	\poU & ~=~ -\,C\sqrt{\xi}\; \p_r K_0\left(\frac{2}{N}\mUm\,r\right),
\\[3mm]
	\plN & ~=~ C\,\mNm\,\sqrt{\xi}\; K_0\left(\mNm\, r\right),
&
	\poN & ~=~ -\,C\sqrt{\xi}\; \p_r K_0\left(\mNm\,r\right)\,,
\end{align*}
	with an arbitrary constant $ C $.
	At larger $ r $, but still in the region where $ K_0 $ does not fall off exponentially,
	the asymptotics of the above solution is
\beq
\label{ttail}
	\poU ~~=~~ \poN ~~\simeq~~ C\,\frac{\sqrt{\xi}}{r}~.
\eeq

	Despite the fact that two equations in \eqref{free_strans} are independent,
	there is only one undetermined constant $ C $.
	This fact is not seen at large $ r $, since the U(1) and SU($N$) profile functions got untied
	in this domain.
	To correlate the latter functions, we will have a look at a domain closer to the core of the string and
	justify the equality $ \poU ~=~ \poN $ in \eqref{ttail}.

	{\flushleft{\it Intermediate-$r$ domain: $ r ~\lesssim~ 1/(g\sqrt{\xi}) $}.}\\[2mm]
We proceed with finding only the zero-mode profiles $ \poU $ and $ \poN $, 
sufficient for establishing
	their mutual normalization.
	The functions $ \plU $ and $ \plN $ can be found in a similar fashion. 

	Now we do not drop the gauge functions $ f $ and $ f_N$.
	This seems to be more general than the large-$ r $ case.
	To be able to solve the Dirac equations, all we can do is assume $ \mu $ to be very large.
	The gaugino functions $ \lambda $ can still be found from {\it e.g.} the first and 
	third equations in \eqref{lambdaeqs}.
	However, effectively one can discard them.
	Indeed, Eqs.~\eqref{heavy_gauginos} suggest that $ \lambda_0 $'s must be suppressed as $ O(1/\mu) $
	with respect to $ \psi_1 $ and as $ O(1/\mu^2) $ with respect to $ \psi_0 $, which we temporarily
	accept to be finite at $ 1/\mu \to 0 $.
	Assuming this, and taking the first and third equations in \eqref{psieqs}, one observes
	that to the leading order in $ 1/\mu $ the gaugino contributions can be dropped.

	Taking the sum and the difference of these equation, one arrives at
\begin{align}
\notag
&
	\left\{ \p_r ~+~ \frac{1}{r} ~-~ \frac{1}{Nr}\left(f + (N-1)f_N\right)\right\}
		\lgr \poU  + (N-1) \poN \rgr  ~~=~~ 0 \,,
	\\[2mm]
\label{eqn_tr_interm}
&
	\left\{ \p_r ~+~ \frac{1}{r} ~-~ \frac{1}{Nr}\left(f - f_N\right)\right\}
		\lgr \poU - \poN \rgr ~~=~~ 0~.
\end{align}
	These equations are nothing but the first order equations for the string profiles, {\it cf.} Eqs.~\eqref{foes}, 
	and hence the above linear combinations have to be proportional to $ \phi_1/r $ and $ \phi_2/r $, 
	respectively.
	However, whereas $ \phi_1(r) $ vanishes at the origin as $ O(r) $, and $ \phi_1/r $ is well defined, 
	$ \phi_2(r) $ is nonvanishing at zero, and $ \phi_2(r)/r $ is divergent at the origin.
	Demanding finiteness of the zero modes at the string core, one concludes that the second linear combination
	in \eqref{eqn_tr_interm} must vanish everywhere, {\it i.e.} $ \poU ~=~ \poN $.
	Therefore, 
\[
	\poU ~~=~~ \poN ~~\propto~~ \frac{\phi_1}{r}~.
\]
	This agrees with the analysis at large $ r $ above, which states that they need to be proportional
	to $ \sqrt{\xi}/r $ at large $ r $.
	We have proved that there is a connection between the U(1) quark modes and the SU($N$) modes, and that 
	there is only one arbitrary constant $ C $.
	This constant can always be absorbed into the normalization of the supertranslational modes.
	We fix it for convenience as
\begin{equation}
\label{tail_strans}
	\poU ~~=~~ \poN ~~\equiv~~ \frac{2}{N}\, \frac{\phi_1}{r}~, \qquad\qquad  C ~=~ \frac{2}{N}~.
\end{equation}

\subsubsection{Superorientational zero modes}
	We deal in a similar fashion with the superorientational modes.
	Resolving the gauginos from the first and  third equations of \eqref{fermeqs},
\begin{align}
\notag
	\lambda_+ & ~~=~~ \frac{i}{\sqrt{2}\phi_1} \lgr \p_r \psi_+ ~-~ \frac{1}{Nr}(f - f_N)\psi_+ \rgr  \,,\\[2mm]
\notag
	\lambda_- & ~~=~~ \frac{i}{\sqrt{2}\phi_2} \lgr \p_r \psi_- ~+~ \frac{1}{r} \psi_- ~-~ \frac{1}{Nr}(f + (N-1)f_N)\psi_- \rgr
\end{align}
	and substituting them into the two remaining equations in \eqref{fermeqs}, while dropping 
	the kinetic terms, one obtains
\begin{align}
\notag
	& \p_r \psi_+ ~-~ \frac{1}{Nr} (f - f_N)\, \psi_+ ~+~ \mNm \frac{\phi_1\phi_2}{\xi}\, \psi_- ~~=~~ 0 \,,\\[2mm]
\label{orient_large_mu}
	& \p_r \psi_- ~+~ \frac{1}{r}\psi_- ~-~ \frac{1}{Nr}(f + (N-1)f_N)\, \psi_- ~+~ \mNm \frac{\phi_1\phi_2}{\xi}\, \psi_+ ~~=~~ 0~.
\end{align}
	We again solve these equations in two domains.
	This analysis is completely similar to that of the translational modes, yet even simpler,
	and we do not give as much detail.

	{\flushleft{\it Large-$r$ domain: $ r \gg 1/(g\sqrt{\xi}) $}.}\\[2mm]
	At large $ r $ we put the string profiles to their vacuum values, {\it i.e.} $ f = f_N $ $ = 0 $,
	and $ \phi_{1,2} = \sqrt{\xi} $, and solve for $ \psi_- $,
\[
	\psi_- ~~=~~ -\,\frac{1}{\mNm} \p_r \psi_+~,
\]
	Thus, we  have
\[	
	\p_r^2\, \psi_+  ~~+~~ \frac{1}{r}\p_r\, \psi_+ ~-~ (\mNm)^2\, \psi_+ ~~=~~ 0~.
\]
	The latter equation then yields 
\beq
\label{sorient_K}
	\psi_+ ~~=~~ \mNm K_0(\mNm r)~, \qquad\qquad \psi_- ~~=~~ -\, \p_r K_0(\mNm r)~,
\eeq
	with just one arbitrary constant which we have implicitly put to unity.

	{\flushleft{\it Intermediate-$r$ domain: $ r ~\lesssim~ 1/(g\sqrt{\xi}) $}.}\\[2mm]
In this domain we do not put the string profiles to their asymptotic values, but instead
	take $ \mu $ to be very large.
	Then, ignoring the small mass terms in \eqref{orient_large_mu} we get
\begin{align*}
	& \p_r \psi_+ ~-~ \frac{1}{Nr}(f - f_N)\, \psi_+ ~~=~~ 0\,, \\[2mm]
	& \p_r \psi_- ~+~ \frac{1}{r}\psi_- ~-~ \frac{1}{Nr}(f + (N-1)f_N)\, \psi_- ~~=~~0~.
\end{align*}
	These equations again remind us the string profile equations \eqref{foes}, and we write
\[
	\psi_+ ~~=~~ c_1 \phi_2 \qquad\qquad	\psi_- ~~=~~ c_2 \frac{\phi_1}{r}~.
\]
	Neither of $ c_1 $ and $ c_2 $ have to vanish now.
	On the other hand we know from the analysis at large $ r $ that there cannot be two independent
	constants.
	In fact, since we have fixed the normalization of the zero modes at large $ r $, Eq.~\eqref{sorient_K}, 
	we have no more freedom, and can determine $ c_1 $ and $ c_2 $ by matching the large-$r$
	and medium-$r$ solutions at $ r = 1/m_W $,
\begin{equation}
\label{tail_sorient}
	\psi_+ ~~=~~ \mNm \frac{\ln m_W/\mNm}{\sqrt{\xi}} \phi_2~,
	\qquad\qquad
	\psi_- ~~=~~ \frac{1}{\sqrt{\xi}} \frac{\phi_1}{r}~,
	\qquad\qquad
	m_W ~~=~~ g_2\sqrt{\xi}~.
\end{equation}
	
	Equations \eqref{tail_strans} and \eqref{tail_sorient} explicate the long-range tails of the
	right-handed zero modes.
	One observes that in the limit $ \mu \to \infty $ the latter become non-nor\-ma\-li\-za\-ble, the fact
	related to the presence of the Higgs branch in \none SQCD, to which our theory flows.

\section{Bifermionic Coupling}
\label{BC}
\setcounter{equation}{0}

	The bifermionic coupling, given in the fourth line of Eq.~\eqref{world02}, arises from the 
	kinetic terms of the superorientational and supertranslational moduli.
	On the one hand, it can be found from the microscopic theory. 
	To detect its presence in the effective action, one only has to substitute the corresponding zero modes 
	into the kinetic terms of our theory \eqref{fermact}.
	On the other hand, its magnitude determines the deformation parameter  of the world sheet theory, 
\[
	\mathcal{W}_{1+1} ~~=~~ \frac{1}{2}\,\delta\,\Sigma^2~,
	\qquad\qquad
	\tgamma ~~=~~ \frac { \sqrt{2}\,\delta } { \sqrt{ 1 +  2 |\delta|^2 } }\,,
\]
	and ultimately gives the connection sought for, between the microscopic deformation parameter $ \mu $, and 
	the macroscopic ``response'' $ \delta $.
	In \cite{SYhet} it was shown that in the SU(2) theory the dependence of one on another is logarithmic at large 
	$ \mu $.
	More generally, this fact should not depend on $ N $, and now we will show that it does not.

	We substitute the zero modes obtained earlier in this section into the kinetic terms of the theory
	(\ref{fermact}).
	The anticipated result is the kinetic part of the effective sigma model, with the following structure:
\begin{align}
\label{world_unnorm}
	S_{\rm 1+1}^{{\rm CP}(N-1)} ~~=~~  \int d^2x 
	\Biggl\lgroup\; 
	&
		2\pi\xi \lgr   \frac{1}{2} \left(\p_k \vec{x}_0 \right)^2
				~+~  \frac{1}{2} \ov{\zeta}{}_L\, i\p_R \zeta_L 
				~+~  \frac{I_\zeta}{2} \ov{\zeta}{}_R\, i\p_L \zeta_R
			\rgr
	\\[3mm]
\notag
	~~+~~  
	&\;
	2\beta \lgr \left|\p_k n \right|^2  ~+~ \left(\ov{n}\p_k n\right)^2  
		~+~ \ov{\xi}{}_L\, i\p_R \xi_L  ~+~  I_\xi\, \ov{\xi}{}_R\, i\p_L  \xi_R
		\right .
	\\[3mm]
\notag
	&\;
		\left. 
		~+~ I_{\zeta\xi} 
			\left(  i\p_L\ov{n}\, \xi_R \zeta_R ~+~  \ov{\xi}{}_R \, i\p_L n \zeta_R \right)
	 \rgr
	~~+~~  \text{four-ferm.int.}
	\Biggr\rgroup ~.
\end{align}
	The constants  $ I_\zeta $ and $ I_\xi $ are the normalizations
	of the corresponding kinetic terms which will be determined upon the substitution of the
	zero modes.
	The integrals $ I_\zeta $ and $ I_\xi $ are expected to be dependent on $ \mu $ as they
	normalize the right-handed moduli, which are affected by the \none deformation
	under consideration.

	The strength of the bifermionic coupling $ I_{\zeta\xi} $ is  our prime  interest.
	In \eqref{world_unnorm} we  implicitly assumed that $ \mu $ is real, and therefore
	the induced $ \tgamma $ is also expected to be real.
	Using the zero-mode expressions \eqref{ftprofile} and \eqref{fprofile} we calculate this coupling,
\begin{align}
\notag
	\mc{L}_{\rm 1+1} ~~\supset~~
	&
	\frac{2\pi}{g_2^2} \times  
	4 \lgr  i\p_L \ov{n}\, \xi_R \, \zeta_R  ~+~  \ov{\xi}{}_R \, i \p_L n\, \ov{\zeta}{}_R \rgr
	\\[2mm]
\notag
	&
	\times
	\int r\, dr 
	\biggl\lgroup  ( \rho(r)-1 ) \left( \loN \lambda_-   ~+~   \llN \lambda_+ \right)  \\[2mm]
\label{biferm_general}
	&
	\qquad\qquad~~
	+\,g_2^2 \frac{N}{4} \left( \poN \psi_-   ~+~   \plN \psi_+ \right)   
	\\[2mm]
\notag
	&
	\qquad\qquad~~
	+\, g_2^2 \frac{\rho(r)}{4} \Bigl\{ \left(\poU ~-~ \poN\right) \psi_-   \\[2mm]
\notag
	&
	\qquad\qquad\qquad\;\;
	     		         	-~ \left(\plU ~+~ (N-1)\plN\right) \psi_+ \Bigr\} 
	\biggr\rgroup~.
\end{align}
	We can handle this expression in the limits of small or large $ \mu $.

	For small $ \mu $, we again accept a convenient relation, Eq.~\eqref{mueq}
\[
	g_1^2 \sqrt{\frac{N}{2}}\, \mu_1 ~~=~~ g_2^2 \mu_2~,
\]
	and substitute the profile functions \eqref{tzeroorder} -- \eqref{firstorder} into Eq.~\eqref{biferm_general},
\begin{align}
\label{biferm_small}
	&
	\frac{2\pi}{g_2^2} \times  
	4 \lgr  i\p_L \ov{n}\, \xi_R \, \zeta_R  ~+~  \ov{\xi}{}_R \, i \p_L n\, \ov{\zeta}{}_R \rgr
	\left( -\, \frac{\mu_2 g_2^2}{2\sqrt{2}} \right)
	\\[4mm]
\notag
	&
	\times
	\int r dr 
	\lgr  \frac{g_2^2(\phi_1^2 - \phi_2^2)^2}{\phi_2^2} 
			\left( 1 \;+\; \frac{1}{N} f \;+\; \frac{2N - 1}{N}f_N \right) 
			\;+\;
		4 g_2^2 (\phi_1^2 \;-\; \phi_2^2) f_N \rgr,
\end{align}
	to the first order in $ \mu $.
	Since the normalization constants of the right-handed modes $ I_\zeta $, $ I_\xi $ do not
	depend on $ \mu $ in the leading order (recall, they are \ntwo supersymmetric in the leading order),
	all dependence of $ \tgamma $ and $ \delta $ on $ \mu $ is given by \eqref{biferm_small},
	and it is linear.

	The large-$\mu$ limit is the most interesting limit, as in this limit the relation between $ \delta $ 
	and $ \mu $ ceases to be linear. 
	To be able to compare $ I_{\zeta\xi} $ to the coefficient in front of the bifermionic mixing term in 
	the CP($ N-1 $) model \eqref{world02}, one needs to properly normalize the kinetic terms in \eqref{world_unnorm}.
	In other words, one needs the expressions for all  $ I_\zeta $, $ I_\xi $ and $ I_{\zeta\xi} $ in terms
	of the zero-mode profiles.
	The contributions of the heavy gaugino, however, now are discarded, and we obtain
\begin{align}
\notag
	I_\zeta & ~=~ \frac{N}{2\xi} 
		\int r dr 
			\lgr \left(\poU\right)^2 \;+\; \left(\plU\right)^2 \;+\; 
				(N-1)\left\{\left(\poN\right)^2 \;+\;
				            \left(\plN\right)^2 \right\} 
			\rgr \!, \\[3mm]
\label{normint}
	I_\xi & ~=~ 2g_2^2\, 
		\int r dr \lgr  \psi_+^2 \;+\; \psi_-^2 \rgr\!, \\[3mm]
\notag
	I_{\zeta\xi} & ~=~
		\frac{N}{2}\, g_2^2\, 
		\int r dr \lgr \poN\, \psi_- \;+\; \plN\, \psi_+ \rgr \!,
\end{align}
	where we also omitted the gauge field contribution from \eqref{biferm_general} since it would not
	produce large logarithms which we are after. 

	Substituting the long-range tails \eqref{tail_strans} and \eqref{tail_sorient} into \eqref{normint}
	we arrive at
\begin{align}
\notag
	I_\zeta & ~~=~~ 2\, \ln \frac{m_W}{m_L} ~+~ O(1)~, \\[2mm]
\label{logs}
	I_\xi & ~~=~~ 2\,g_2^2 \, \ln \frac{m_W}{m_L} ~+~ O(1)~, \\[2mm]
\notag
	I_{\zeta\xi} & ~~=~~ g_2^2\sqrt{\xi}\, \ln \frac{m_W}{m_L} ~+~ O(1)~, 
\end{align}
	where
\[
	m_L ~~=~~ \frac{\xi}{\mu_2}~,
\]
	and the integration is carried out over the range $ 1/m_W \lesssim r \lesssim 1/m_L $.

	Normalizing the fields $ \xi $ and $ \zeta $ canonically (modulo the factor  $ 2\beta $, 
	{\it cf.} Eq.~\eqref{world02}) and comparing the result for $ I_{\zeta\xi} $ with \eqref{world02},
	we conclude that
\beq
\label{gammaresult}
	\tgamma ~~=~~ 
		\frac { \sqrt{2}\,\delta } { \sqrt{ 1 + 2 | \delta |^2 } } 
		~~=~~ 1 ~~+~~ O\left(\frac{1}{\ln\left(\frac{g_2^2\mu}{m_W}\right)}\right)~
\eeq
	(remind that in this paper $ \tgamma $ is related to $ \gamma $ of \cite{SYhet} as
	$ \tgamma = \sqrt{2/\beta}\, \gamma $).
	This leads to the following result:
\beq
\label{deltaresult}
	\delta~~=~~ 
	{\rm const} \cdot \sqrt{\ln\, \frac{g_2^2\mu}{m_W}}~,
	\qquad\qquad \text{as $\mu ~\to~ \infty$}~.
\eeq
	The latter exhibits a nonlinear dependence of the world sheet deformation 
	parameter $ \delta $ on the 
	supersymmetry breaking parameter $ \mu $, independently of the number of colors $ N $.

\section{Physics of the heterotic  CP({\boldmath $N-1$}) model}
\setcounter{equation}{0}
\label{largeN}

In this section we will briefly review physics of the heterotic \ntwoo  CP$(N-1)$
model which, as was showed above, is an effective low-energy theory on the world sheet 
of the non-Abelian string in the deformed \ntwo SQCD. The
 \ntwoo  CP$(N-1)$ model was solved in \cite{SYhetlN} in the  large-$N$ approximation.
This method was suggested by Witten  \cite{W79} and used to solve nonsupersymmetric
and \ntwot supersymmetric CP$(N-1)$ model. The results obtained in 
\cite{SYhetlN} unambiguously demonstrate that the  \ntwoo supersymmetry is in fact spontaneously broken
(see also \cite{Tohetdyn}) and the vacuum energy density does not vanish. This means
that the elementary string tensions  are lifted from their classical BPS values,
\beq
T~~=~~ 2\pi\xi ~+~ \frac{N}{4\pi}\,\Lambda_{\sigma}(1-e^{-u}),
\label{vacenergy}
\eeq
where the deformation parameter $u$ is related to $\delta$ as follows:
\beq
u ~~\equiv~~ \frac{16\pi\beta}{N}\,|\delta|^2\,.
\label{u}
\eeq
At large $\mu$, the  parameter $u$ behaves as 
\beq
u ~~\sim~~ \ln{\frac{g_2^2\mu}{m_W}}\,.
\eeq

Just as its \ntwot cousine, the heterotic \ntwoo model has $N$ degenerate vacua.
They are labeled by a nonvanishing VEV of the complex scalar $\sigma$ (the former \ntwot
superpartner of the photon $A_k$). Namely, $\sigma$ is given by
\beq
\sigma ~~=~~  \frac1{\sqrt{2}}\,\exp{\left(-\,\frac{u}{2}~+~\frac{2\pi i k}{N}\right)},
\qquad k=0,...,(N-1).
\label{sigmavev}
\eeq
Thus, indeed,  we have $N$ vacua  in the world sheet model, {\it i.e.} $N$ degenerate
elementary non-Abelian strings in the bulk theory. In fact, the field $\sigma$  is 
proportional
to the bifermion condensate $\bar{\xi}_L\xi_R$ and its VEV signals  the 
chiral symmetry breaking. The point is that classically the model at hand has the U(1) chiral symmetry which is, however, broken 
by the chiral anomaly down to a discrete $Z_{2N}$ symmetry. The nonvanishing VEV
(\ref{sigmavev})  breaks this $Z_{2N}$ symmetry further down to $Z_2$.

As soon as we have $N$ vacua in the model we also have kinks interpolating between
these vacua. From the standpoint  of the bulk theory they are interpreted as confined 
monopoles \cite{T,SYmon,HT2,SYnone,GSYmmodel,SYhetlN}. These monopoles are free to move
along the string. This should be contrasted to the nonsupersymmetric CP$(N-1)$ model where
kinks  and anti-kinks form bound states, kink-anti-kink
``mesons'' \cite{W79}. From the four-dimensional point of view this means that the
monopoles (in addition to four-dimensional confinement) are in the phase of  two-dimensional
confinement which ensures that monopole and anti-monopole  form a 
meson-like bound state on the string \cite{GSY05}.

The presence of $N$ degenerate vacua (\ref{sigmavev}) shows that the monopoles
are deconfined in the two-dimensional sense (on the string) in the theory at hand.

	When we tend $ \mu $  to infinity, all  $N$ vacua (\ref{sigmavev}) coalesce and 
the world sheet theory supposedly flows to
 a conformal phase \cite{SYhetlN}.
	However, this theory cannot be trusted in this limit because of the presence
	of the massless modes \eqref{light} which make the string swell;  higher-order 
corrections on
	the world sheet become increasingly important.

%
%
\section{Conclusion}
\setcounter{equation}{0}

This paper concludes the program started in \cite{SYhet},
namely direct derivation of the world sheet theory for heterotic non-Abelian strings
starting from the bulk theories with \ntwo supersymmetry broken down to \none
by the mass term of the adjoint fields. 
If in the previous work the bulk theory analyzed was U(2)
supersymmetric QCD, now we analyzed U$(N)$ with arbitrary $N$.
To this end we had to explicitly obtain all fermion zero modes.
We managed to accomplish this task in two limits: small and large $\mu $.
An explicit relation between the parameters of the world sheet and bulk theories
was found.  
Our results for U$(N)$ are very similar to those for U(2).

\section*{Acknowledgments}

The work of PAB was supported in part by the NSF Grant No. PHY-0554660. PAB is grateful for kind
hospitality to FTPI, University of Minnesota, where a part of this work was done. 
The work of MS was supported in part by DOE grant DE-FG02-94ER408. 
The work of AY was  supported 
by  FTPI, University of Minnesota, 
by RFBR Grant No. 09-02-00457a 
and by Russian State Grant for 
Scientific Schools RSGSS-11242003.2.

\newpage
\appendix
\section*{Appendix: Notation}
%
%

\setcounter{equation}{0}

Throughout the paper we use the matrix notation for the matter fields
\[
	q^{kA}~, \qquad \psi^{kA}~,\qquad \ov{q}_{Ak}~, \qquad \ov{\psi}_{Ak}~,
	\qquad k~=~1,...,N, \qquad A~=~1,...,N,
\]
which allows us to conveniently treat them on the same footing with the gauge fields, when the gauge
symmetry is broken, and parametrize all of them as SU($N$) matrices. 
The trace needs to be performed when the action is built.

The gauginos $ \lambda^f $ carry the SU(2)$_R$ index $ f $, with $ \lambda^1 $ belonging
to the \none gauge supermultiplet, and $ \lambda^2 $ being part of the adjoint supermultiplet. 
For the barred gauginos $ \ov{\lambda}_f $ the same placement is defined for the lower
index $ f $.
These SU(2)$_R$ indices are raised and lowered by the SU(2) metric tensor
\[
	\epsilon_{fg} ~=~ 
			\lgr \begin{matrix}
			     	\ 0\  &  \ 1\   \\
				 -1\  &  \ 0\  
			     \end{matrix} \rgr,
	\qquad 
	\epsilon^{fg} ~=~ 
			\lgr \begin{matrix}
				\ 0\ &   -1\   \\
				\ 1\ &  \ 0\ 
			     \end{matrix} \rgr.
\]

We use euclidean space-time throughout the paper, both in the four-dimensional and two-dimensional
cases.
As for the four-dimensional spinors, their indices are raised and lowered by the SU(2) metric tensor,
similarly to the above,
\[
	\psi_\alpha ~=~ \epsilon_{\alpha\beta}\, \psi^\beta, \qquad
	\ov{\psi}{}_{\dot{\alpha}} ~=~ \epsilon_{\dot{\alpha}\dot{\beta}}\, \ov{\psi}{}^{\dot\beta}, \qquad 
	\psi^\alpha ~=~ \epsilon^{\alpha\beta}\, \psi_\beta, \qquad
	\ov{\psi}{}^{\dot{\alpha}} ~=~ \epsilon^{\dot\alpha\dot\beta}\, \ov{\psi}{}_{\dot\beta}~,
\]
	where
\[
	\epsilon_{\alpha\beta} ~=~ \epsilon_{\dot\alpha\dot\beta} ~=~
			\lgr \begin{matrix}
			     	\ 0\  &  \ 1\   \\
				 -1\  &  \ 0\  
			     \end{matrix} \rgr,
	\qquad \text{and} \qquad
	\epsilon^{\alpha\beta} ~=~ \epsilon^{\dot\alpha\dot\beta} ~=~
			\lgr \begin{matrix}
				\ 0\ &   -1\   \\
				\ 1\ &  \ 0\ 
			     \end{matrix} \rgr.
\]
The contractions of the spinor indices are short-handed as
\[
	\lambda\psi ~=~ \lambda_\alpha\, \psi^\alpha\,, \qquad
	\ov{\lambda\psi} ~=~  \ov{\lambda}{}^{\dot\alpha}\, \ov{\psi}{}_{\dot\alpha}\,.
\]
The sigma matrices for the euclidean space we take as
\[
	\sigma^{\alpha\dot\alpha}_\mu ~=~  \lgr {\bf 1},\quad -i\,\tau^k \rgr^{\alpha\dot{\alpha}}
	\hspace{-2.0ex},
	\qquad
	\ov{\sigma}{}_{\dot\alpha\alpha\, \mu} ~=~ 
			\lgr {\bf 1},\quad i\, \tau^k \rgr_{\dot\alpha\alpha},
\]
where $ \tau^k $ are the Pauli matrices.

Reduction to two dimensions can be conveniently done by picking out $ x^0 $ and $ x^3 $ 
as the world sheet (or ``longitudinal'') coordinates, and integrating over the orthogonal coordinates. 
One then identifies the lower-index spinors as the two-dimensional left- and right-handed chiral spinors
\[
	\xi_{R} ~=~ \xi_{1}\,, \quad\qquad
	\xi_{L} ~=~ \xi_{2}\,, \quad\qquad\qquad
	\ov{\xi}{}_{R} ~=~ \ov{\xi}{}_{\dot{1}}\,, \quad\qquad
	\ov{\xi}{}_{L} ~=~ \ov{\xi}{}_{\dot{2}}\,.
\]

In the world sheet theory, the CP($N-1$) indices are written as upper ones for the orientational variables
\[
	n^l\,, \quad \xi^l\,,
\]
and lower ones for the conjugate moduli
\[
	\ov{n}{}_l\,, \quad \ov{\xi}{}_l\,, 
\]
where $ l~=~1,\, ...,\, N $.
In the geometric formulation of CP($N-1$), global indices are written upstairs in both cases, only
for the conjugate variables the indices with bars are used 
\[
	\phi^i\,,\ \psi^i\,, \qquad \ov{\phi}{}^\bi\,,\ \ov{\psi}{}^\bi\,, 
	\qquad\qquad i,\ \bi ~=~ 1,\,...,\,N-1\,,
\]
and the metric $ g_{i\bj} $ is used to contract them.

\vspace{1cm}

%
%

\end{document}